\documentclass[11pt, dvips]{article}

\RequirePackage[OT1]{fontenc}
\RequirePackage{amsthm,amsmath}
\RequirePackage[numbers]{natbib}
\RequirePackage[colorlinks]{hyperref}
\RequirePackage{hypernat}

\usepackage{graphicx}
\usepackage{amssymb}

\newtheorem{theorem}{Theorem}[section]
\newtheorem{proposition}{Proposition}[section]
\newtheorem{definition}{Definition}[section]
\newtheorem{remark}{Remark}[section]

\newcommand{\br}{\begin{remark}\rm}
\newcommand{\er}{\end{remark}}

\def\diag{\mathop{\rm diag}}
\def\sspan{\mathop{\rm span}}

\def\be{\begin{equation}}
\def\ee{\end{equation}}
\def\bea{\begin{eqnarray*}}
\def\eea{\end{eqnarray*}}

\def\last#1{{\underline{#1}}}
\def\first#1{{\mathstrut\overline{#1}}}

\newcommand{\ontop}[2]{\genfrac{}{}{0pt}{0}{#1}{#2}}
\newcommand{\noc}[1]{\small{\emph{#1}}}

\def\suml{\sum\limits}

\begin{document}

\title{On the choice of parameters in Singular Spectrum Analysis and related subspace-based methods}
\author{Nina Golyandina\footnote{nina@gistatgroup.com, Department of Mathematics, St.Petersburg State University,
Universitetsky pr. 28, 198504, St.Petersburg, Russia}}
\date{}

\maketitle

\begin{abstract}
In the present paper we investigate methods related to both the
Singular Spectrum Analysis (SSA) and subspace-based methods in signal processing. We describe common and specific features
of these methods and  consider different kinds of problems
solved by them such as signal reconstruction, forecasting and parameter estimation.
General recommendations on the choice of parameters to obtain
minimal errors are provided. We demonstrate that the optimal choice depends on the particular problem.
For the basic model `signal + residual' we show that the error behavior
depends on the type of residuals, deterministic or stochastic, and whether the noise is white or red.
The structure of errors and the convergence rate are also discussed.
The analysis is based on  known theoretical results and  extensive computer simulations.
\end{abstract}

%\begin{keyword}[class=AMS]
%\kwd[Primary ]{62M20}
%\kwd{62F10}
%\kwd{62F12}
%\kwd[; secondary ]{60G35}
%\kwd{65C20}
%\kwd{62G05}
%\end{keyword}


\section{Introduction}

In the 1970--80s many papers describing methods based on the Singular Value Decomposition (SVD) of
a specially constructed Hankel matrix
have been published,
see \cite{Kung.etal1983,
Kumaresan.Tufts1982, Basilevsky.Hum1979,Broomhead.King1986,Fraedrich1986, Cadzow1988} among many others.
Many modern methods are based on the ideas proposed in these papers and these methods have proved to be
very useful in many applied areas. The list of the methods includes
HSVD (Hankel SVD, \cite{Barkhuijsen.etal1987}) and
 HTLS (Hankel Total Least Squares, \cite{VanHuffel.etal1994}) which are mostly used in
 the nuclear magnetic resonance spectroscopy,
ESPRIT  \cite {Roy.Kailath1989} used in the Direction-Of-Arrival problems, and Singular Spectrum Analysis (SSA) which
has been applied in many different areas.
%in various fields, such as [13] in climatology, etc.

All these methods have several common features (up to using the same algorithms),
but each of them also has some special features.
The purpose of this paper
is to show some commonalities and specifics of these  methods and to analyze the optimal choice of parameters.
Since  SSA is not limited to any particular area of application, we mostly base our investigation on SSA.

Let us briefly describe Basic SSA  \cite{Golyandina.etal2001} for the real-valued case.
Let us observe a one-dimensional time series $F_N=(f_0,\ldots,f_{N-1})$ of length $N$.
We suppose that $F_N$ is a sum  of several unknown but identifiable components
and we are interested in some of them, for example, trend or regular oscillations.

%Henceforth, the transition to the complex-valued case can be performed by taking a matrix's
%conjugate transpose instead of its transpose.

\smallskip\noindent
\textbf{Embedding\ } Given a {\em window length} $L$ ($1 < L < N$) one proceeds with constructing
$K=N-L+1$ lagged ($L$-lagged) vectors
$X_i=(f_{i-1},\ldots,f_{i+L-2})^\mathrm{T}$, $1\leq i\leq K$,
 and composing them into the matrix $\mathbf{X}=[X_1:\ldots:X_K]$,
which is called the {\em $L$-trajectory matrix}. Note that $\mathbf{X}$ is a Hankel matrix.% with entries $x_{ij}=f_{i+j-2}$.

\smallskip\noindent
\textbf{Decomposition\ } The key step in SSA is the SVD of the trajectory matrix:
\be
\label{eq:SVD}
    \mathbf{X} = \suml_{i=1}^d \sqrt{\lambda_i}U_i V_i^\mathrm{T},
\ee
where $\lambda_1\geq \ldots \geq \lambda_d>0$ are the positive eigenvalues of the matrix
$\mathbf{C}=\mathbf{X} \mathbf{X}^\mathrm{T}$ arranged in the nonincreasing order,
$\{U_1,\ldots,U_d\}$ is the orthonormal system of the corresponding eigenvectors of the matrix  $\mathbf{C}$ ($U_i$ are left singular vectors of $\mathbf{X}$),
and $V_i=\mathbf{X}^\mathrm{T} U_i/\sqrt{\lambda_i}$, $i=1,\ldots,d$, are the factor vectors (right singular vectors).
Note that $U_1,\ldots,U_d$ form an orthonormal basis of the column space $\mathcal{L}_d$ of $\mathbf{X}$, which is
called the \emph{trajectory space}, $\mathcal{L}_d=\sspan(U_1,\ldots,U_d)$. The triple $(\sqrt{\lambda_i}, U_i, V_i)$ is called the $i$-th
\emph{eigentriple} (or ET) and $\sqrt{\lambda_i}$ is called the $i$-th singular value of $\mathbf{X}$.

\smallskip\noindent
\textbf{Grouping\ } After {\em grouping  the eigentriples} by choosing a partition
of the set $\{1,\ldots,d\}$ onto $m$ disjoint subsets $I_1, \ldots, I_m$,
one obtains the decomposition
\bea
\mathbf{X}=\mathbf{X}_{I_1}+\ldots+\mathbf{X}_{I_m},
\eea
where $\mathbf{X}_I = \suml_{i\in I} \sqrt{\lambda_i}U_i V_i^\mathrm{T}$.
Alternatively, we can consider the grouping step as a decomposition of the trajectory space into the
orthogonal sum of subspaces: $\mathcal{L}_d=\bigoplus\limits_{k=1}^m \mathcal{L}^{(k)}$, where
$\mathcal{L}^{(k)} = \sspan(U_i, i\in I_k)$.

\smallskip\noindent
\textbf{Reconstruction (diagonal averaging)\ } One can {\em reconstruct}
the components of the original series
by the {\em diagonal averaging} of each component $\mathbf{X}_{I_k}$:
\bea
F_N = {\widetilde F}^{(1)} + \ldots + {\widetilde F}^{(m)}\,,
\eea
where  ${\widetilde F}^{(k)}=({\widetilde f}^{(k)}_0,\ldots,{\widetilde f}^{(k)}_{N-1})$
and ${\widetilde f}^{(k)}_i$ is  the average value along the {\em $i$-th secondary diagonal} of $\mathbf{X}_{I_k}$.

\medskip
Although the described algorithm is designed to decompose the original time series into a sum of an arbitrary number of components,
we will be considering the problem of decomposition into a sum of two time series (shortly, t.s.) components.
The problem of decomposition into several components can be reduced to the sequential extraction
of the components one by one (e.g., first by
extracting the trend with a small window length and then by
extracting the periodicity from
the residual using a large window length, see \cite{Golyandina.etal2001}).
In addition, a considerable number of methods of time series analysis (including ESPRIT examined in this paper)
solve the problem of analyzing noisy signals and, in fact, consider only two t.s. components:
signal and noise.

Thus, suppose that we observe $F_N=S_N+R_N$, where $S_N$ is the component of interest (e.g., signal) and
$R_N$ is the residual component. The residual $R_N$ can be either a random noise or a deterministic component,
or a mixture of both.

The estimator $\widetilde{S}_N={\widetilde F}^{(1)}$  of the time series $S_N$ (so called
\emph{reconstructed} signal) can be considered as the main result of application of SSA.
Appropriate reconstruction of the signal can be obtained in the case of its approximate separability from the residual.
Typical examples of pairs of approximately separable t.s. components (see \cite{Golyandina.etal2001} for details)
are trend (i.e., a slowly-varying component) and noise,
trend and cyclic components, and two cyclic components with different frequencies.
In general, we can talk about approximate separability
of one of the following t.s. components: trend, cycle or noise,  from a mix of the other ones.

Suppose that the t.s. component $S_N$, which we are interested in, satisfies  the following two properties.
First, let $S_N$ be deterministic.
Second, let $S_N$ be leading, i.e. $S_N$ generates the $r$ leading SVD components in the decomposition
\eqref{eq:SVD} with indices from $I=\{1,\ldots,r\}$.
In fact, we suppose that the minimal singular value of the trajectory matrix
$\mathbf{S}$ of $S_N$ is larger or asymptotically larger than the maximal singular value of the trajectory matrix $\mathbf{R}$ of $R_N$.

The deterministic structure of $S_N$ and its approximate separability imply that $S_N$ can be approximated
by a time series of finite $L$-rank \cite{Golyandina.etal2001}. This means that in
\eqref{eq:SVD} all except the $r$ leading eigenvalues are close to zero.
In this paper we investigate the asymptotic properties as the time series length $N$ tends to infinity.
We consider an infinite time series $F=(f_0,\ldots,f_n,\ldots)$ and analyze
the finite-length time series $F_N$ consisting of its first $N$ terms.
For an infinite time series $S$, the following assertion is valid: $S$ \emph{has finite rank} $r$ (its $L$-trajectory space $\mathcal{L}_r^{(\mathrm{s})}$
has dimension $r$ for any $L\geq r$) iff $S$ \emph{is governed by a linear recurrent formula} (LRF) \cite{Golyandina.etal2001}.
For finite-rank subseries $S_N$ the SVD of $L$-trajectory matrix $\mathbf{S}$ has exactly $r$ positive singular values and
$\mathbf{S}$ is rank deficient. If $S_N$ is interpreted as
a signal, then its trajectory space $\mathcal{L}_r^{(\mathrm{s})}$ is called the \emph{signal subspace}.

Note that if a time series is governed by an LRF, then it
 can be represented as a sum of products of polynomials, exponentials and sinusoids,
and in this case it makes sense to use a parametric setup for the problem.
On the other hand,
the SSA algorithm of signal reconstruction is essentially nonparametric and can be applied to a class
of time series which is wider than the set of perturbed finite-rank signals.

Statements of problems within the framework of SSA can differ in the following aspects.

\begin{itemize}
\item[A1]
\label{item:features}
Features of interest: we can be interested either in the signal $S_N$ as a whole
or in some of its characteristics.
In particular, if $S_N$ has finite rank, then it has a known parametric form and
we can be interested in parameter estimation. The most elaborated problem is the
estimation of damping factors and frequencies of exponentially damped sinusoids in noise.
For solving this estimation problem it is sufficient to know only  $r$ leading
eigenvectors in \eqref{eq:SVD}, more specifically, the subspace spanned
by these eigenvectors (i.e., the estimated subspace of the signal $S_N$); see
e.g. ESPRIT-like methods.

\item[A2]
\label{item:residual}
Type of residuals: the residual $R_N$ is either deterministic or stochastic (or it has both
random and deterministic components).
These cases correspond to different properties of the SSA decomposition and
cause different characteristics of estimators.
For example, a finite-rank $S_N$ will be the leading component if
the deterministic residual is bounded by some constant, while white noise
can have any variance for large time series lengths $N$ and window lengths $L$.
Also, the structure of the stochastic noise (e.g., white or red) can influence the behavior of estimation errors.

\item[A3]
\label{item:window}
Choice of the window length: either we can vary the window length $L$ or $L$ is fixed.
In the former case, the problem of the optimal choice of $L$ arises.
Then, the asymptotic behavior depends on whether $L$ tends to
infinity as $N\rightarrow \infty$ or not. If we consider a matrix $L\times K$ with a fixed number of rows as an input,
then the only way is to fix $L$ and to consider $K$ tending to infinity.
The other possible reason for choosing a not very large $L$ is the
computational cost. However, there are recent computational advances, which make calculations very fast, see
\cite{Korobeynikov2009}.
\end{itemize}

%\smallskip
In this paper, we consider different statements of the problem within Aspect
A1 and then analyze errors and parameter choice rules following Aspects
A2 and A3.

The main information about the time series structure that we obtain within the framework of SSA
is contained in the set of eigentriples $(\sqrt{\lambda_i},U_i,V_i)$. Consequently we obtain
not only the reconstructed signal but also much additional information
about $S_N$. In addition to the problem of reconstructing $S_N$, we consider the problems of
signal forecasting and signal parameter estimation.

Note that the SVD is determined only by the set of eigenvectors $U_i$,
since $\lambda_i=\|\mathbf{X}^\mathrm{T} U_i\|^2$ and $V_i=\mathbf{X}^\mathrm{T} U_i/\sqrt{\lambda_i}$,
where $\mathbf{X}$ is the trajectory matrix of the observed time series. Consequently,
$\widetilde{\mathbf{S}}=\sum_{i=1}^r U_i U_i^\mathrm{T}\mathbf{X}=\mathrm{P}_r\mathbf{X}$,
where $\mathrm{P}_r$ is the orthogonal projection on $\mathcal{L}_r$.
Therefore, we can say that the set of eigenvectors (along  with the time series $F_N$)
completely determines the whole SVD expansion and, therefore, the results of forecasting and
parameter estimation.
Thus it is natural to start the investigation with the estimation of eigenvectors
 or, equivalently, the eigenspace estimation. Note that the problem of
estimation of the factor vectors $V_i$ becomes the problem
of estimation of the eigenvectors $U_i$ by changing the window length from $L$ to $N-L+1$.

Let us remark that the transformations applied to the eigenvectors in order to obtain, for example, the frequency estimates
can tremendously change the structure of estimation errors.
Nevertheless, in Section~\ref{sec:projector} we consider the errors of signal subspace estimation as
the starting point of the investigation.
Sections \ref{sec:reconstruction} and \ref{sec:forecast} contain the results on reconstruction
and forecasting based on the chosen subspace.
Section~\ref{sec:estimation} is devoted to parameter estimation within the framework of the subspace-based methods
of signal processing including ESPRIT. In Sections \ref{sec:projector}--\ref{sec:estimation} we give general recommendations
on the choice of the window length $L$ which are based on simulations and known theoretical results.
In Section~\ref{sec:convergence} we consider the convergence rate in different conditions (a fixed window length $L$
or a window length proportional to the time series length) for the problems investigated in previous sections.

Let us remark that the results of Sections~\ref{sec:projector}--\ref{sec:convergence} are valid under the condition of
strong separability of the signal from the residual. Section \ref{sec:noseparability} deals with several examples,
in which there is no strong separability.
In Section~\ref{sec:stationary} we consider several versions of Basic SSA and demonstrate some examples of application
of the versions designed for the stationary time series to non-stationary ones.

In Section~\ref{sec:origins} we briefly describe some origins of the SVD providing
the key step of the SSA algorithm.
We are interested in these origins, since they imply different views on the problem statement and
on the parameter choice.

\section{Signal subspace}
\label{sec:projector}
We will generally rely on the results of the paper \cite{Nekrutkin2010},
which is devoted to the discussion of convergence and also contains
the main error terms and their upper bounds.

As a measure of the error for the subspace approximation we consider the spectral norm of the difference
of projectors on the true subspace and the estimated subspace. Note that this norm is equal
to the sine of the largest principal angle between these subspaces.
The aim of this section is to investigate the dependence of the approximation error on the window length.

Let $S_N$ be a signal of rank $r$.
By $\mathrm{P}_r^{(\mathrm{s})}$
we denote the orthogonal projector on the signal subspace $\mathcal{L}_r^{(\mathrm{s})}$, which is spanned
by the left singular vectors $U_1^{(\mathrm{s})},\ldots,U_r^{(\mathrm{s})}$ of the signal trajectory matrix $\mathbf{S}$,
and by $\mathrm{P}_r$ we denote
 the orthogonal projector on the estimated signal subspace $\mathcal{L}_r=\sspan(U_1,\ldots, U_r)$,
where $U_1,\ldots, U_r$ are the $r$ leading left singular vectors
of the trajectory matrix of the observed time series $F_N$.
Note that we can easily calculate the estimation error $\|\mathrm{P}_r^{(\mathrm{s})}-\mathrm{P}_r\|$, since
the cosine of the largest principal angle between $\mathcal{L}_r^{(\mathrm{s})}$ and $\mathcal{L}_r$ is equal to
the $r$-th  eigenvalue of the matrix $\mathbf{U}_r^{(\mathrm{s})}\mathbf{U}_r^\mathrm{T}$, where
$\mathbf{U}_r^{(\mathrm{s})}=[U_1^{(\mathrm{s})}:\ldots:U_r^{(\mathrm{s})}]$ and
$\mathbf{U}_r=[U_1:\ldots:U_r]$ (see e.g. \cite[p.~18]{Bjorck1996}).

Let us consider five examples of time series $f_n=s_n+r_n$, $n=1,\ldots,N$,
\begin{align}
\label{eq:1-1}
s_n= &\  1,\ r_n=-c(-1)^n,\\
\label{eq:det}
s_n= &\  b^n\cos(2\pi n/10),\ r_n=c,\\
\label{eq:white}
s_n= &\  b^n\cos(2\pi n/10),\ r_n=\sigma\varepsilon_n,\\
\label{eq:mix}
s_n= &\  b^n\cos(2\pi n/10),\ r_n=(\sigma\varepsilon_n+c)/\sqrt{2},\\
\label{eq:red}
s_n= &\  b^n\cos(2\pi n/10),\ r_n=\sigma\eta_n.
\end{align}
Here $\varepsilon_n$ is a white gaussian noise with variance $1$ and $\eta_n$ is the autoregressive process of order 1
(red noise) with parameter $\alpha$ and variance $\mathbf{D}\eta_n=1$, that is, $\eta_n=\alpha\eta_{n-1}+\epsilon_n$, where $\epsilon_n$
has variance  $1-\alpha^2$. In this section, we set $c=\sigma=0.1$, $\alpha=0.5$, $b=1$, and $N=100$.

We choose the level of noise in the time series \eqref{eq:det}--\eqref{eq:red} to have the same signal-to-noise ratio (SNR), which
is conventionally determined as the ratio of the average of squared signal values to the average of squared residual values
(or to the variance of residuals if they are random).

Generally, the SNR  does not determine the size of the errors of estimates obtained by the SSA processing,
since the SNR does not take into consideration the time series length.
In fact, the SNR  can be used to compare
the quality of processing of time series of equal lengths. However, we cannot say that SSA separates signal and noise
only if the SNR is larger than a specific value; for example, for any small SNR
a sine-wave signal is asymptotically separated from a white noise as $N \rightarrow \infty$ and $L\sim  \beta N$, $0<\beta<1$.

The time series~\eqref{eq:1-1} is included, since we can compare the results with those in \cite[Section 4.2.1]{Nekrutkin2010}.
It appears that the main term of perturbation found in \cite{Nekrutkin2010} is almost equal to
the whole error. Moreover, the behavior of errors depends on whether the lengths of the window and the time series are even or odd.
In addition, this is an example
of a time series that produces the projector error having the first-order term with respect to the perturbation level
which is not the main term of the error as $N\rightarrow \infty$.

The time series \eqref{eq:det}--\eqref{eq:mix} differ in the structure of residuals: deterministic, random, or combined.
The time series \eqref{eq:red} is used to consider a noise that differs from the white noise.

In the case of random residuals, we compute either MSD (mean square deviation) or RMSE (square root of mean square error) as a measure of accuracy.
Generally, these criteria yield very similar results. The difference between them is that we can compute
the square root before averaging (MSD) or after averaging (RMSE) for the simulation results.
In the examples of this section we estimate MSD using 100 simulations.

\begin{center}
             \begin{figure}[h!]
                    \includegraphics[height = 55mm]{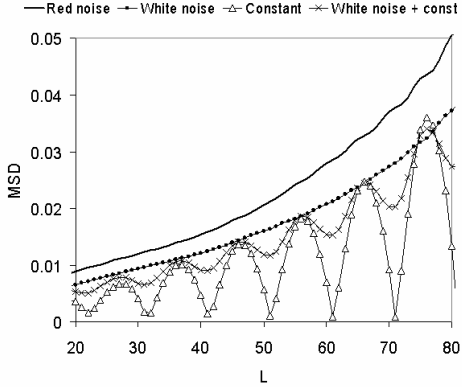}
                    \vspace{-3mm}\caption[t]{MSD of projector estimates: different types of residuals, t.s.~\eqref{eq:det}--\eqref{eq:red}}
                    \label{fig:cf_resP}
             \end{figure}
\end{center}

We present the results of simulation study for the time series~\eqref{eq:1-1}--\eqref{eq:red}
in Figures~\ref{fig:sawP}--\ref{fig:cf_resP}.
Figures \ref{fig:sawP} and \ref{fig:cos_constPlog} show the errors in estimation of the projector
$\mathrm{P}_r^{(\mathrm{s})}$ on the signal subspace for the examples with deterministic residuals.
One can see a tendency for the errors to increase as the window length increases.
However, for window lengths $L$ that are divisible by periods of the time series components, errors
generally decrease. This reflects the influence of the multiplicity of $L$ and/or $K$ to the
periods of time series components. Note that if both $L$ and $K$ are divisible by the periods
(by 2 in the first example and by 10 in the second one), then the projector perturbation is equal to 0. This corresponds to the case of
bi-orthogonality of the trajectory matrices of $S_N$ and $R_N$ and therefore to the case  of exact separability.
If only $L$ (or $K$) is divisible by the period, then this case
can be called left (or right) orthogonality.
Thus, if the residual is deterministic and $S_N$ contains a periodic component,
then we observe two effects: the specific behavior of errors in the case
of window lengths divisible by  the period and a periodic behavior of errors in the general case.

It is clear that if the residual $R_N$ contains noise, then the exact orthogonality cannot be achieved.
Figures \ref{fig:cos_noisePlog} (the logarithmic scale) and \ref{fig:cos_noiseP} (the original scale)
demonstrate that the decrease of errors for special window lengths does not occur.

In the case of combined residuals (Fig.~\ref{fig:cos_noiseconstPlog}), the behavior of
projector errors  inherits the properties of errors for both pure random and deterministic residuals.
Below we show that this feature is valid for other kinds of problems.

To show that the fact that the noise is  red (rather than white)  does not interfere with the extraction of the signal,
we consider the time series~\eqref{eq:red} with a red noise,
see Fig.~\ref{fig:cos_rednoisePlog}. Fig.~\ref{fig:cf_resP} compares MSD for different structures
of residuals (recall that SNR is the same). One can see that a red noise yields a slightly worse accuracy.
Errors for the time series \eqref{eq:mix} lie between those for the time series \eqref{eq:det} and the time series \eqref{eq:white}.

To summarize, in this section we demonstrated the behavior of errors of projector estimates for different
types of residuals. However, we are usually interested in certain features of the signal space, rather than in the projector itself.
Therefore, the results of this section provide just the basic information, which can be used for explanation of
further results.

\begin{center}
             \begin{figure}[h!]
                    \includegraphics[height = 57mm]{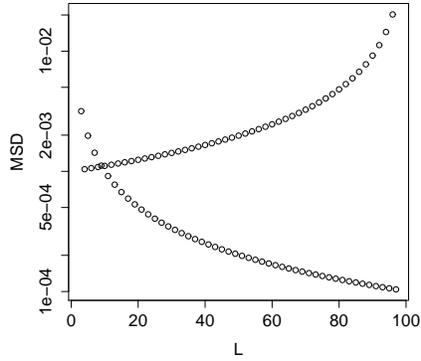}
                    \vspace{-3mm}\caption[t]{MSD of projector estimates: deterministic residuals, t.s.~\eqref{eq:1-1}
                    (log-scale)}
                    \label{fig:sawP}
             \end{figure}
             \begin{figure}[h!]
                    \includegraphics[height = 57mm]{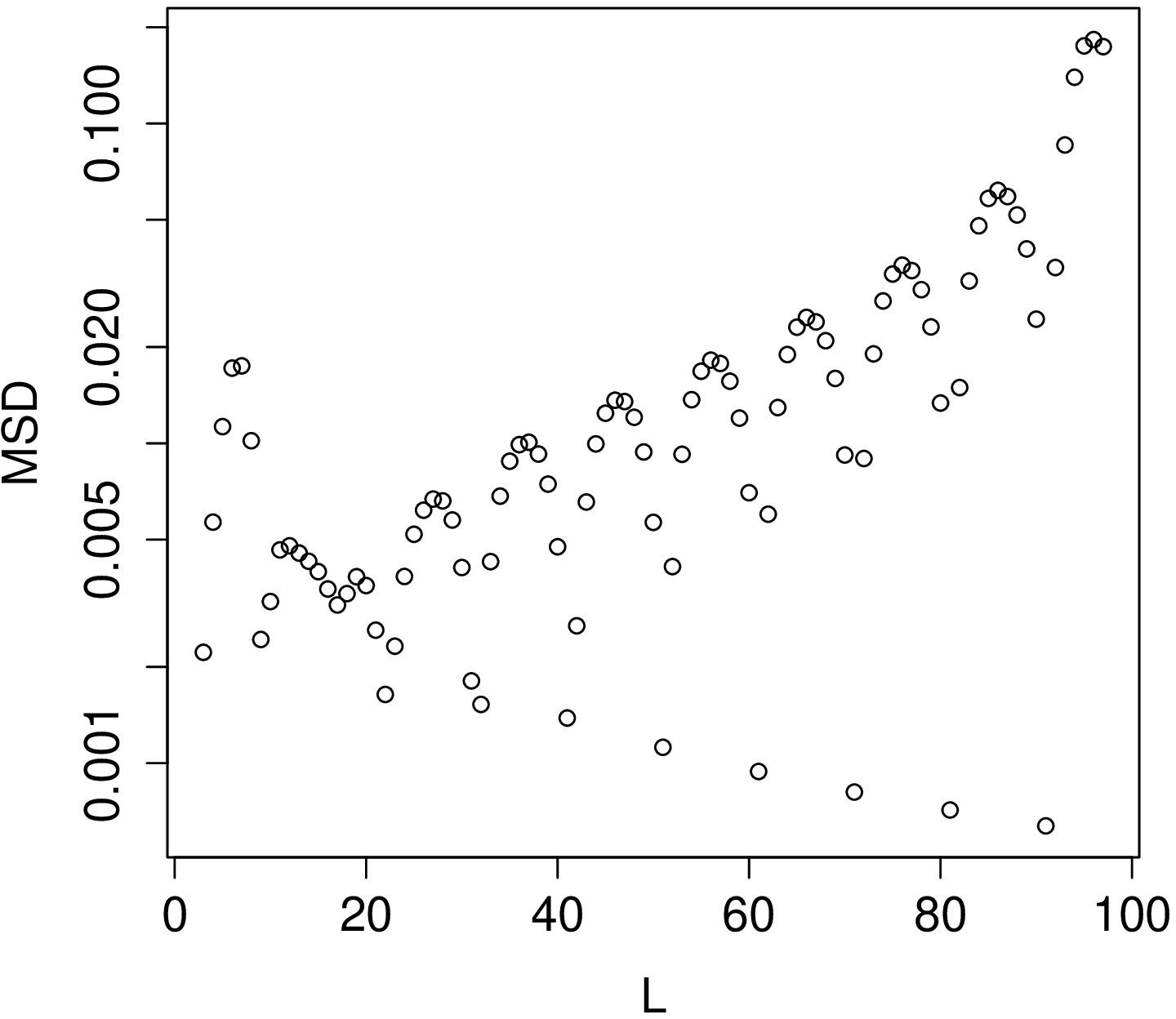}
                    \vspace{-3mm}\caption[t]{MSD of projector estimates: deterministic residuals, t.s.~\eqref{eq:det}
                    (log-scale)}
                    \label{fig:cos_constPlog}
             \end{figure}
\end{center}

\begin{center}
             \begin{figure}[h!]
                    \includegraphics[height = 57mm]{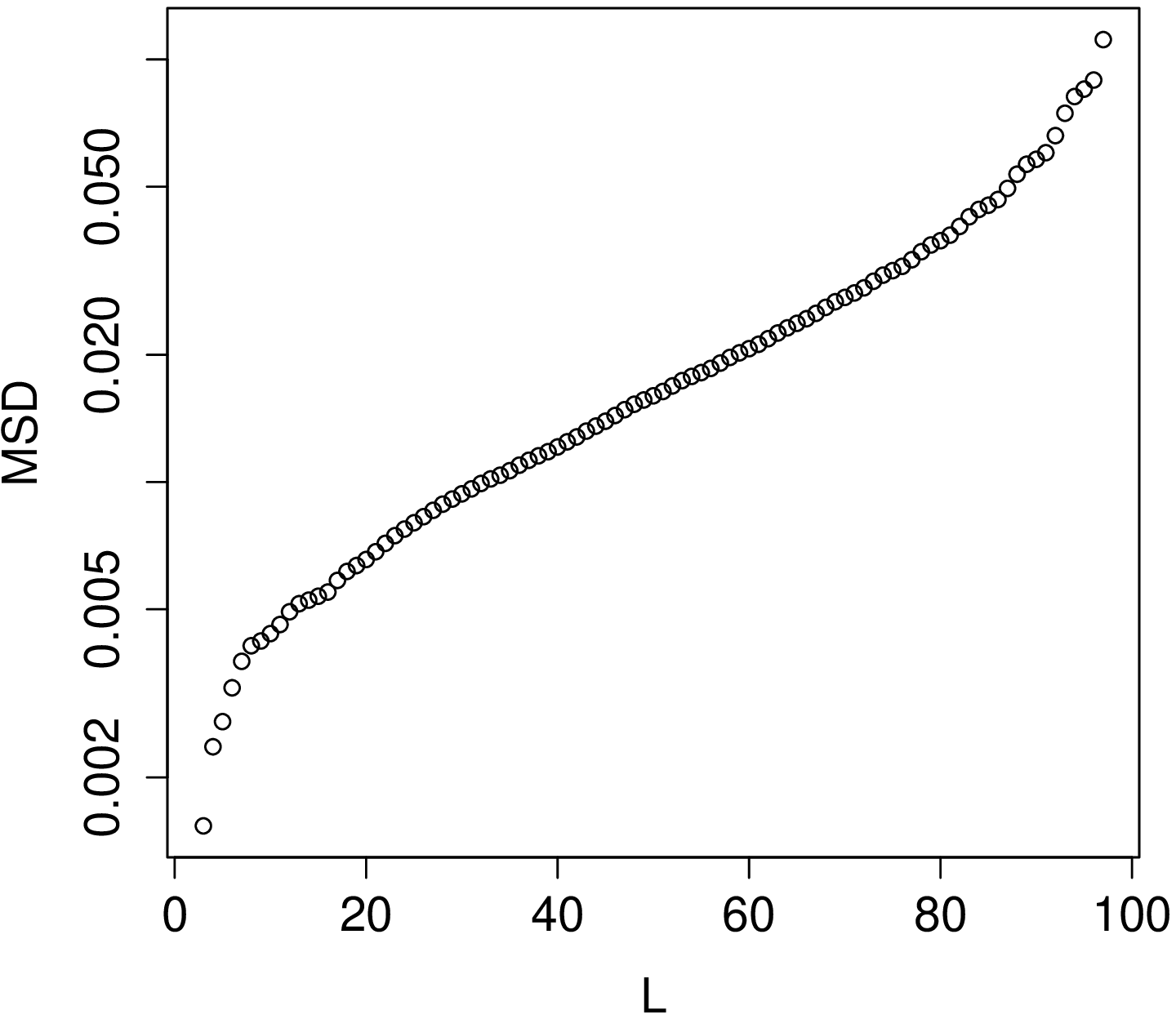}
                    \vspace{-3mm}\caption[t]{MSD of projector estimates: white-noise residuals, t.s.~\eqref{eq:white}
                    (log-scale)}
                    \label{fig:cos_noisePlog}
             \end{figure}
             \begin{figure}[h!]
                    \includegraphics[height = 57mm]{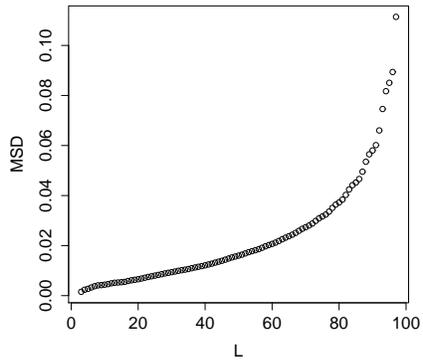}
                    \vspace{-3mm}\caption[t]{MSD of projector estimates: white-noise residuals, t.s.~\eqref{eq:white} (original scale)
                    }
                    \label{fig:cos_noiseP}
             \end{figure}
\end{center}

\begin{center}
             \begin{figure}[h!]
                    \includegraphics[height = 57mm]{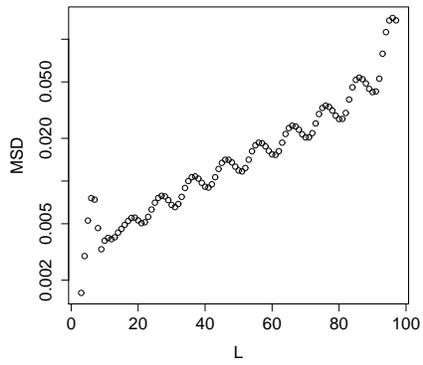}
                    \vspace{-3mm}\caption[t]{MSD of projector estimates: mixed residuals, t.s.~\eqref{eq:mix}
                    (log-scale)}
                    \label{fig:cos_noiseconstPlog}
             \end{figure}
             \begin{figure}[h!]
                    \includegraphics[height = 57mm]{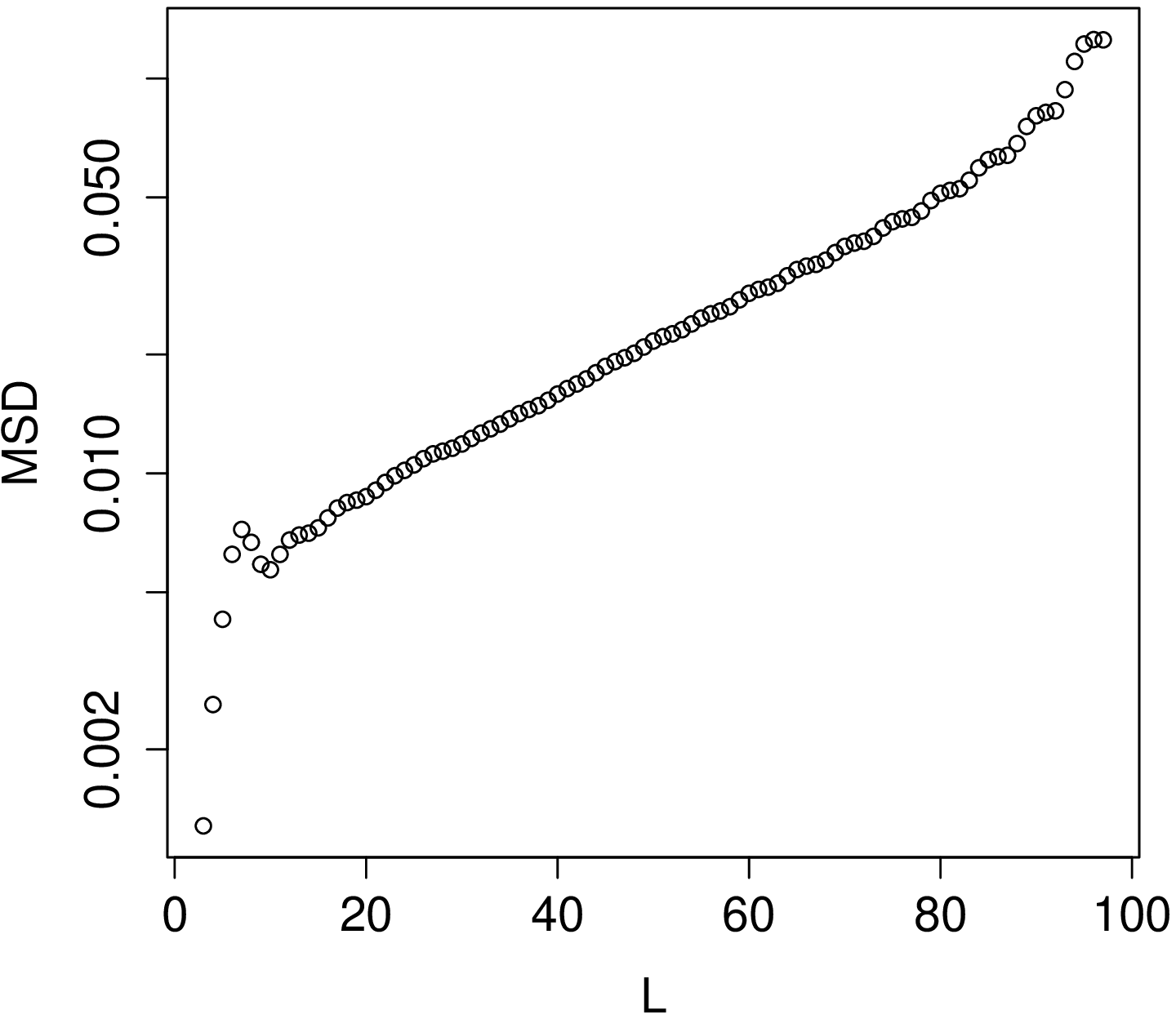}
                    \vspace{-3mm}\caption[t]{MSD of projector estimates: red-noise residuals, t.s.~\eqref{eq:red}
                    (log-scale)}
                    \label{fig:cos_rednoisePlog}
             \end{figure}
\end{center}

\section{Signal extraction}
\label{sec:reconstruction}

Recall that the reconstructed signal is obtained by applying the diagonal averaging to
the reconstructed matrix calculated by the formula
$\widetilde{{\mathbf{S}}}={\mathbf{U}}_r \mathbf{\Lambda}_r^{1/2} {\mathbf{V}}_r^\mathrm{T}$, where
$\mathbf{U}_r=[U_1:\ldots:U_r]$,  $\mathbf{V}_r=[V_1:\ldots:V_r]$, and
$\mathbf{\Lambda}_r=\diag(\lambda_1,\ldots,\lambda_r)$; $U_i$, $V_i$ and $\lambda_i$ are defined in
\eqref{eq:SVD}.
Note that the columns of the matrix ${\mathbf{U}}_r$ form a basis of
the perturbed signal subspace for the window length $L$ while the columns of
${\mathbf{V}}_r$ form a basis of the perturbed signal subspace
for the window length $N-L+1$. First, this means that the results are the
same for the window lengths $L$ and $N-L+1$.
Then, as is shown in Figures \ref{fig:sawP}--\ref{fig:cf_resP}, the signal-subspace perturbation grows
as $L$ increases and decreases  as $N-L+1$ increases. Thus, the resultant errors are caused by
these contradictory tendencies. Fig.~\ref{fig:cos_noiseP} demonstrates that
the growth rate of errors in the projector estimation is larger  for large $L>N/2$. Therefore, there is no surprise
that the reconstruction errors are large for small window lengths due to large errors in $\mathbf{V}_r$.
However, the question about the optimal window lengths remains open.

Further we consider the dependence of the reconstruction error $\widetilde{s}_l-s_l$ on the window length in several examples.
Note that an explicit asymptotic form is known for the example of noisy  constant signal with
$s_n\equiv c$ \cite{Golyandina.Vlassieva2009}. The paper \cite{Golyandina.Vlassieva2009}
contains an explicit expression for the variance of the first-order reconstruction errors, where
 the first order is given with respect to the perturbation
of the signal by $R_N$ and  not with respect to $N$.
Strictly speaking, it is not necessary for the first-order error to be the main term
of the error as $N\rightarrow \infty$.
It has been checked by computer simulations that in the case of pure random noise
we can consider the first-order error as the main error term (it is not true in the general case,
see \cite{Nekrutkin2010}).
Computer simulations confirm that the qualitative results for the constant signal are valid
for many other types of signals as well, including oscillations. To describe these results let us
present the formula for the dependence of the asymptotic errors on the window length
for a constant signal \cite{Golyandina.Vlassieva2009}.

Let the window length $L \sim \beta N$, $0 < \beta \le 1/2$, and $l$ be the index of the time series point,
$l\sim \gamma N/2$, $0\le \gamma\le 1$, as $N\rightarrow \infty$. The value $\gamma=1$ corresponds to the middle of the time series;
consequently, we present the formula for the first half of the time series with window lengths smaller
than one-half of the time series length. Then the variance of the first-order errors has the following asymptotic form:
\begin{gather}
\label{eq:MSSA}
    \mathbb{D}s_l^{(1)} \sim \frac{\sigma^2}{N}
        \begin{cases}
            D_1(\beta, \gamma),\ 0\le \gamma \le 2\min(\beta,1-2\beta), \cr
            D_2(\beta, \gamma),\ 2\min(\beta,1-2\beta) < \gamma < 2\beta, \cr
            D_3(\beta, \gamma),\ 2\beta \le \gamma \le 1,
        \end{cases}
\end{gather}
as $N\rightarrow \infty$, where
\begin{eqnarray*}
    D_1(\beta, \gamma) &=& \frac{1}{12\beta^2(1-\beta)^{2}} \Big({\gamma}^{2}\left(1+\beta\right)-2\gamma\beta\left(1+\beta\right)^2\\
    &+&  4\beta\left(3-3\beta+2{\beta}^{2}\right)\Big),\\
    D_2(\beta, \gamma) &=&  \frac{1}{6\beta^2\left(1-\beta\right)^{2}{\gamma}^{2}} \Big({\gamma}^{4}+2\gamma^3\left(3\beta-2-3\beta^2 \right) \nonumber\\
    &+&\left.2\gamma^2\left(3-9\beta+12\beta^2-4\beta^3\right) \right.    \\
    &+&4\gamma\left(-1+4\beta-3\beta^2-4\beta^3+4\beta^4\right) \nonumber \\
    &+&\left(8\,\beta-56\,{\beta}^{2}+144\,{\beta}^{3}
    -160\,{\beta}^{4}+64\,{\beta}^{5}\right)\Big)\label{eq:DL2LLessKformula},\\
    D_3(\beta, \gamma) &=& \frac{2}{3\beta}\label{eq:DL3LLessKformula}.
\end{eqnarray*}
The points of change between the cases in \eqref{eq:MSSA} correspond to $l=K-L$
(i.e., $\gamma=2(1-2\beta)$) and $l=L$ ($\gamma=2\beta$).
The former  point of change is present if $K<2L$ ($\beta>1/3$).
Note that formula~\eqref{eq:MSSA} can be extended to the window lengths $2<L<N-1$ ($0 < \beta < 1$) and
to the indices of time series points $0\le l\le N-1$ ($0\le \gamma\le 2$)
due to the symmetry of errors with respect to the middle of the time series and by the equivalence of
results under the substitution of $L$ for $K$  ($\beta \leftrightarrow 1-\beta$).

When we solve the problem of minimizing RMSE of estimation of $s_l$ at a fixed point $l$, the
optimal window length varies from $N/3$ to $N/2$, see \cite{Golyandina.Vlassieva2009}. This means that even in the case
with a constant signal the optimal window length, which minimizes the reconstruction errors as a whole,
depends on the importance (weights) of each point of the time series. In any case, the general recommendation is to
choose the window length slightly less than one-half of the time series length $N$.
Note that the optimal window length provides a considerable improvement in
the error rate (with respect to the choice $L=N/2$) at the edge time series points,
that is, for $l/N \approx 0$.

\begin{center}
             \begin{figure}[h!]
                    \includegraphics[height = 55mm]{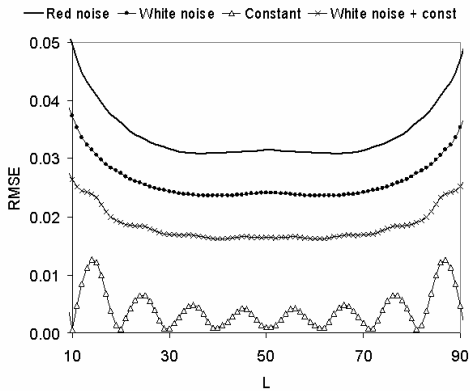}
                    \vspace{-3mm}\caption[t]{RMSE of signal estimates: different types of residuals, t.s.~\eqref{eq:det}--\eqref{eq:red}}
                    \label{fig:cf_resMSE}
             \end{figure}
\end{center}

It has been shown for the projectors in Section~\ref{sec:projector} that for a noisy sine-wave signal (i.e. if the residuals do not contain
a deterministic component), the  divisibility of the window length by the sine-wave period
is not an important issue. The presence of a deterministic component in the residual
makes this divisibility important. A similar effect takes place for the reconstruction errors $\|\widetilde{S}_N-S_N\|$,
see Fig.~\ref{fig:cf_resMSE} for the time series \eqref{eq:det}--\eqref{eq:red}
with the same parameters.

\begin{center}
             \begin{figure}[h!]
                    \includegraphics[height = 55mm]{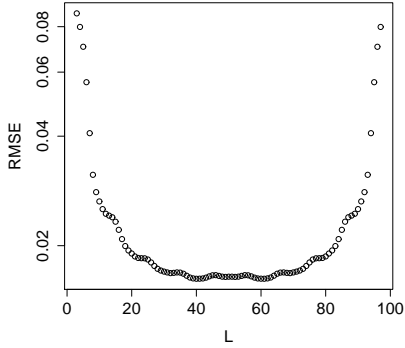}
                    \vspace{-3mm}\caption[t]{RMSE of signal estimates: mixed residuals, t.s.~\eqref{eq:mix}}
                    \label{fig:cos_noiseconstMSElog}
             \end{figure}
             \begin{figure}[h!]
                    \includegraphics[height = 55mm]{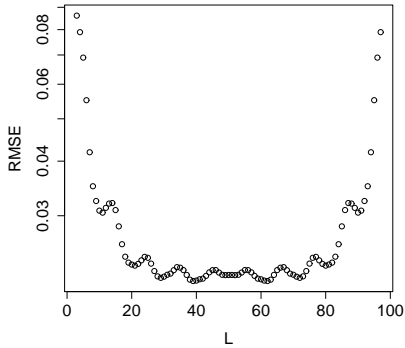}
                    \vspace{-3mm}\caption[t]{RMSE of signal estimates for the last 10 points: mixed residuals, t.s.~\eqref{eq:mix}}
                    \label{fig:cos_noiseconstMSE10log}
            \end{figure}
\end{center}

To study the influence of the window length on RMSE, we consider RMSE for the reconstruction of the ten last points of the signal.
Comparison of Fig. \ref{fig:cos_noiseconstMSElog} and Fig. \ref{fig:cos_noiseconstMSE10log} shows that
the impact of the divisibility  of the window length by the sine-wave period is stronger for the edge points.

In Fig.~\ref{fig:cf_resMSE} we observe that the optimal window length is close to $0.4 N$ in the case of random residual.
However, the divisibility of the window length by the  period (if the residuals contain a deterministic component) can be more important
than the adjustment, e.g., than the transition from $N/2$ to $0.4N$.

\section{Recurrent SSA forecast}
\label{sec:forecast}

\subsection{Theory}
\label{subsec:LRF}
Let us consider the algorithm of recurrent forecasting \cite{Golyandina.etal2001} from the viewpoint
of its connection to the signal subspace estimation. We start with some definitions.

\begin{definition}
A time series $S_N=\{s_i\}_{i=0}^{N-1}$ is \emph{governed by a
linear recurrent formula} (LRF), if there exist  $a_1,\ldots,a_t$ such that
\be
\label{eq:lrf}
s_{i+t}=\sum_{k=1}^t a_k s_{i+t-k},\ 0\leq i<N-t,\ a_t\neq0.
\ee
The number $t$ is called the order of the LRF, $a_1,\ldots,a_t$ are the coefficients of the LRF.
If $t$ is the minimal order of an LRF that governs the time series $S_N$,
then the corresponding LRF is called \emph{minimal}.
\end{definition}

Note that if the minimal LRF governing the signal $S_N$ has order $r$, then $S_N$ has finite rank $r$.

\begin{definition}
A polynomial $P_t(\mu)=\mu^t - \suml_{k=1}^t a_k \mu^{t-k}$ is called a \emph{characteristic polynomial} of
the LRF \eqref{eq:lrf}.
\end{definition}

Let the time series  $S=(s_0,\ldots,s_n,\ldots)$ satisfy the LRF
\eqref{eq:lrf}  with $a_t\neq 0$ and  $i\geq 0$. Consider the
characteristic polynomial of the LRF
\eqref{eq:lrf}
and denote the multiplicities of its (in general, complex)
different roots $\mu_1,\ldots,\mu_p$
by $k_m$, where  $1\leq m\leq p$, $k_1+\ldots+k_p=t$.
Note that the polynomial roots are non-zero as $a_t\neq 0$. Then the following well-known result (see e.g. \cite{Hall1998,Golyandina.etal2001})
provides an explicit form for the series which satisfies the LRF.

\begin{theorem}
\label{th:ANFIN}
The time series  $S=(s_0,\ldots,s_n,\ldots)$ satisfies the LRF $\eqref{eq:lrf}$ for all $i\ge 0$
iff
\be
\label{eq:GEN_REQ}
    s_n = \suml_{m=1}^p \left(\suml_{j=0}^{k_m-1} c_{mj} n^j\right) \mu_m^n,
\ee
where the complex coefficients $c_{mj}$ depend on the first $t$ points
$s_0,\ldots,s_{t-1}$.
\end{theorem}

Note that if the LRF is not minimal, then the  corresponding characteristic polynomial has extraneous
roots. The extraneous roots do not affect the time series behavior, since the coefficients $c_{mj}$ for the corresponding
summands are equal to zero. However, if one applies the LRF to the perturbed initial terms
$\widetilde{s}_0,\ldots,\widetilde{s}_{t-1}$, then the extraneous roots start to affect the forecasting results.
Therefore, the extraneous roots with modules greater than $1$ are the most hazardous, since the extraneous summand $\mu^n$, caused by
an extraneous root $\mu$, $|\mu|>1$, grows to infinity.

Unfortunately, if one analyzes/forecasts a real-life time series, the minimal LRF cannot be estimated with an appropriate accuracy
and hence the presence of extraneous roots should be taken into account.
The estimated LRF can be used both to find a parametric form of the signal (then we should remove extraneous roots)
and also to forecast the time series (then we do not need to know the values of the polynomial roots; however, we would like to have no
extraneous roots beyond the unit circle).

\smallskip\noindent
\textbf{Finding the LRF} Let $P_1\ldots P_r$ be an orthonormal basis of the signal subspace
$\mathcal{L}_r^{(\mathrm{s})}=\sspan\{P_1\ldots P_r\}$ and
$\mathcal{L}_{r+1,L}^{(\mathrm{s})} = \sspan\{P_{r+1}\ldots P_L\}$ be its orthogonal complement.
Denote $A=(a_{L-1},\ldots,a_1,-1)^\mathrm{T}\in \mathcal{L}_{r+1,L}^{(\mathrm{s})}$, $a_{L-1}\neq 0$. Then the time series satisfies the LRF
$s_{n}=\sum_{k=1}^{L-1} a_k s_{n-k}$, $n=L-1,\ldots,N-1$.

Conversely, if a time series is governed by an LRF, then the LRF's coefficients $B=(a_{L-1},\ldots,a_1)^\mathrm{T}$ complemented with $-1$
yield the vector $\left(\ontop{B}{-1} \right)\in \mathcal{L}_{r+1,L}^{(\mathrm{s})}$.
Any LRF that governs the time series can be treated as  a forward linear prediction.
In addition, if we consider a vector from $\mathcal{L}_{r+1,L}^{(\mathrm{s})}$ with $-1$ as the first coordinate,
then we obtain the so-called backward linear prediction \cite{Tufts.Kumaresan1982b}.

Let us denote the matrix $\mathbf{A}$ without the last row by $\last{\mathbf{A}}$ and
the matrix $\mathbf{A}$ without the first row by $\first{\mathbf{A}}$.

From the viewpoint of prediction, the LRF governing a time series of rank $r$ has coefficients derived from the condition
$\last{\mathbf{S}}^\mathrm{T} B = (s_{L-1},\ldots,s_{N-1})^\mathrm{T}$. This system of linear equations
may have several solutions, since the vector $(s_{L-1},\ldots,s_{N-1})^\mathrm{T}$ belongs to the column space
of the matrix $\last{\mathbf{S}}^\mathrm{T}$.
It is well-known that the least-squares solution expressed by the pseudo-inverse to $\last{\mathbf{S}}^\mathrm{T}$
yields the vector $B$ with the minimum norm (the TLS-solution coincides with it).

It can be shown that the minimum-norm solution $B_\mathrm{LS}$ can be expressed as
\be
\label{eq:PVV_F}
B_\mathrm{LS}=(a_{L-1},\ldots,a_1)^\mathrm{T}=\frac{1}{1-\nu^2}\suml_{i=1}^r\pi_i \last{P_i},
\ee
where $\pi_i$ is the last coordinates of $P_i$ and $\nu^2=\sum_{i=1}^r \pi_i^2$.

Thus, one of the vectors from $\mathcal{L}_{r+1,L}^{(\mathrm{s})}$, which equals $A_\mathrm{LS}=
\left(\ontop{B_\mathrm{LS}}{-1} \right)$,
has a special significance and is called the \emph{min-norm (forward) prediction}.
Similarly, we can obtain a formula for the min-norm backward prediction.

It is shown in \cite{Kumaresan.Tufts1980, Golyandina.etal2001} that the forward min-norm prediction vector $A_\mathrm{LS}$
 is the normalized (so that its last coordinate is equal to $-1$)
projection of $\mathbf{e}_L=(\mathbf{0}_{L-1}, 1)^\mathrm{T}$ on the orthogonal complement to the signal
subspace. Therefore, the min-norm prediction vector depends on the signal subspace only.

The paper \cite{Kumaresan.Tufts1983} contains a property of the min-norm LRF, which is very important for forecasting:
all the extraneous roots of the min-norm LRF lie inside the unit circle on the complex plane.
This gives us the hope that in the case of real-life time series (when the min-norm LRF and the related initial data  are perturbed)
the extraneous summands in \eqref{eq:GEN_REQ} decrease and just slightly influence the forecast.
Moreover, in view of results about the distribution of extraneous roots (see \cite{Pakula1987,Usevich2010}),
we can suppose that the extraneous summands are able to compensate one another.

Note that this min-norm LRF forms the basis for the forecasting methods introduced in
\cite{Golyandina.etal2001}. In particular, the recurrent forecast is constructed by applying
formula \eqref{eq:PVV_F} with $P_i=U_i$, where $U_i$ are taken from \eqref{eq:SVD}, to the last $L-1$ reconstructed signal points
$\widetilde{s}_{N-(L-1)},\ldots,\widetilde{s}_{N-1}$.

\subsection{Dependence on the window length}

Let us consider the dependence of the forecast accuracy on the window length $L$ (we consider the signal rank $r$ as given in advance).
Note that the forecasting procedure uses two objects:
the LRF itself and the initial data for this LRF taken from
the last points of the reconstructed signal $\widetilde{S}_N$.
Let us denote the vector constructed from the $L-1$ last signal points by $V$  and
the vector constructed from the $L-1$ last reconstructed (i.e., perturbed) signal points by $V+\Delta V$.
Likewise, we denote the vector of coefficients of the true min-norm LRF by $A$ and the
vector of the estimated LRF coefficients by $A+\Delta A$. Then the forecast error is
$A^\mathrm{T} \Delta V + (\Delta A)^\mathrm{T} V + (\Delta A)^\mathrm{T} \Delta V$.

Therefore, the first-order error consists of two kinds of errors:
\begin{enumerate}
\item the errors in the LRF coefficients that are caused by an error of the projection onto the signal subspace
$(\Delta A)^\mathrm{T} V$;
\item the errors of signal reconstruction $A^\mathrm{T} \Delta V$.
\end{enumerate}

Let us investigate these two error sources separately.
To do this, we apply the LRF that was estimated with the window length $L_{\mathrm{LRF}}$ to the true signal values
and apply the true LRF to the estimated signal values that were reconstructed with the window length $L_{\mathrm{rec}}$.

We consider the time series \eqref{eq:white} with $\sigma=0.1$, $N=399$ and $\ln b=0,\,0.01,\, -0.01$.
Estimation (by 1000 simulations) of RMSE of the one-term ahead forecast is depicted
in Fig.~\ref{fig:for_cos_noise200} for $L_{\mathrm{rec}}=200$. Values of $L_{\mathrm{LRF}}$ are varied from
20 to 380 with increment 20.
Similar graphs for different $L_{\mathrm{rec}}$ are presented in Fig.~\ref{fig:for_cos_noise}. The labels of x-axis
correspond to the values of $L_{\mathrm{rec}}$.

The line marked `total' shows the accuracy (RMSE) of the forecast with the corresponding window lengths.
The line marked `LRF' corresponds to $(\Delta A)^\mathrm{T} V$. The `LRF' errors look like the errors of the estimator $\mathrm{P}_r$
of the projector on the signal space (see Fig.~\ref{fig:cos_noiseP}). This is not surprising since the coefficients of the LRF
are proportional to the projection ${\mathbf{U}}_{r+1,L} {\mathbf{U}}_{r+1,L}^\mathrm{T}\, \mathbf{e}_L={\mathrm{P}}_r^\bot\, \mathbf{e}_L$.
Naturally, the `LRF' errors do not depend on the window length $L_{\mathrm{rec}}$ used for reconstruction.
The typical behavior of this part of the error is as follows: the larger the window length, the larger the error.

\begin{center}
             \begin{figure}[h!]
                    \includegraphics[height = 55mm]{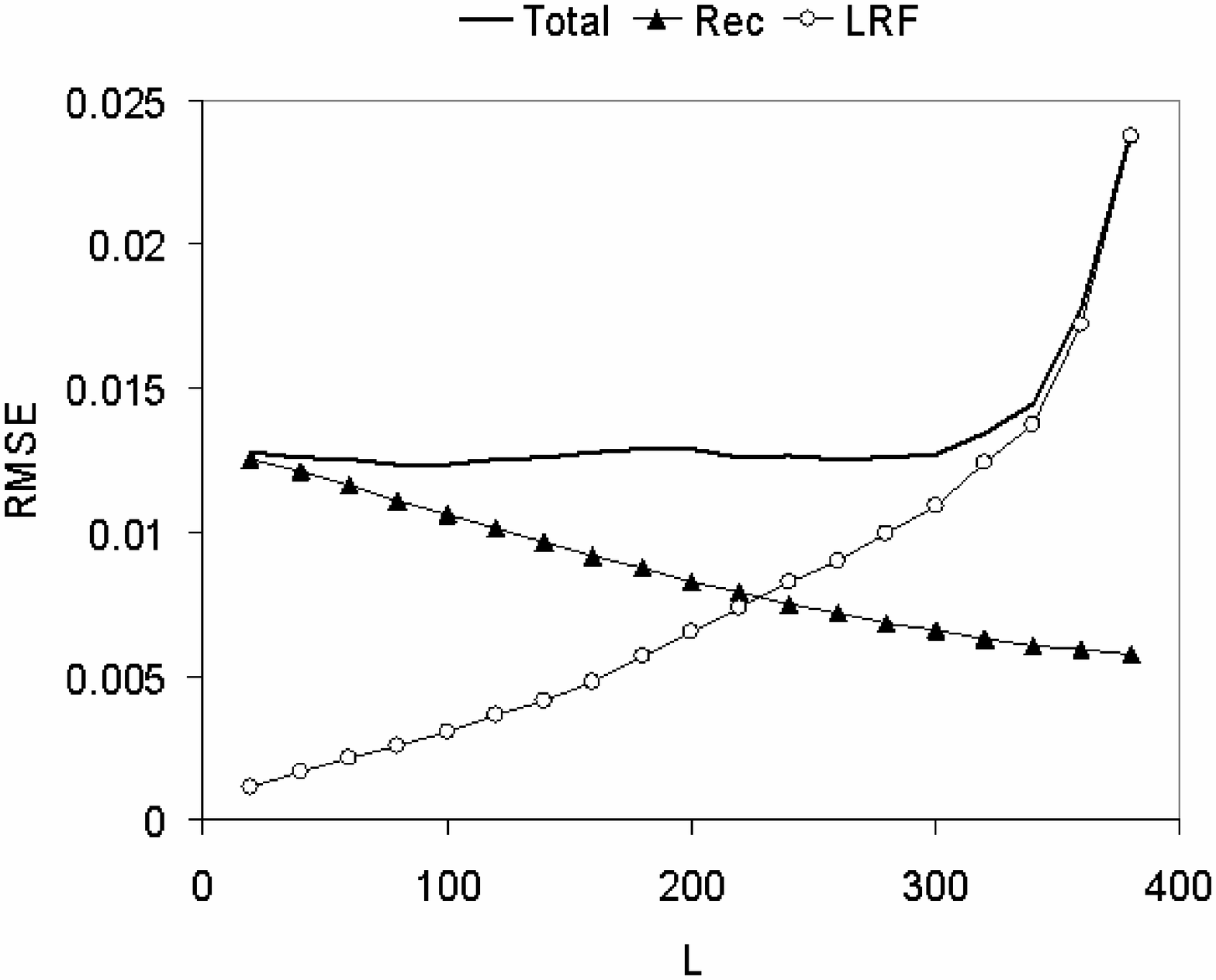}
                    \vspace{-3mm}\caption[t]{RMSE of forecast as a function of $L_{\mathrm{LRF}}$: t.s.~\eqref{eq:white} with $b=1$, $L_{\mathrm{rec}}=200$}
                    \label{fig:for_cos_noise200}
             \end{figure}
             \begin{figure}[h!]
                    \includegraphics[width = 70mm]{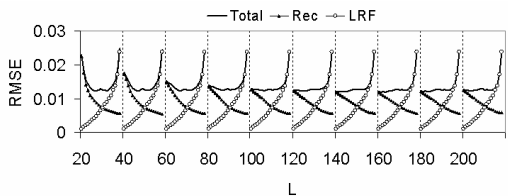}
                    \vspace{-3mm}\caption[t]{RMSE of forecast: t.s.~\eqref{eq:white} with $b=1$,\\ different $L_{\mathrm{rec}}$}
                    \label{fig:for_cos_noise}
             \end{figure}
\end{center}

The line marked `Rec' corresponds to $A^\mathrm{T} \Delta V$.
We can see that the larger the window length, the smaller the error.
This can be interpreted in the following way: extraneous roots that are located closely to the uniform
distribution on a circle compensate one another (see \cite{Pakula1987, Usevich2010} for several results of this kind).

Figures \ref{fig:for_cos_noise200}--\ref{fig:for_cos_noise} show that the accuracy of forecasts is stable within a wide range of window lengths.
In particular, $L_{\mathrm{LRF}}$ and $L_{\mathrm{rec}}$ slightly smaller  than $N/2$ are quite appropriate.
Also, we can take either small $L_{\mathrm{LRF}}$ and $L_{\mathrm{rec}}\sim N/2$ or
$L_{\mathrm{LRF}}\sim N/2$ and small $L_{\mathrm{rec}}$. The former is the preferred choice, since the errors are smaller and
more robust to the changes in the window length, see Fig.~\ref{fig:for_cos_noise}.

Recommendations on the window length choice naturally depend on the forms of the signal and the residual.
Figures~\ref{fig:for_exp1_cos_noise} and \ref{fig:for_exp2_cos_noise} contain RMSE for
damped sine waves. The interpretation of figures and the RMSE behavior are similar.

\begin{center}
             \begin{figure}[h!]
                     \includegraphics[width = 70mm]{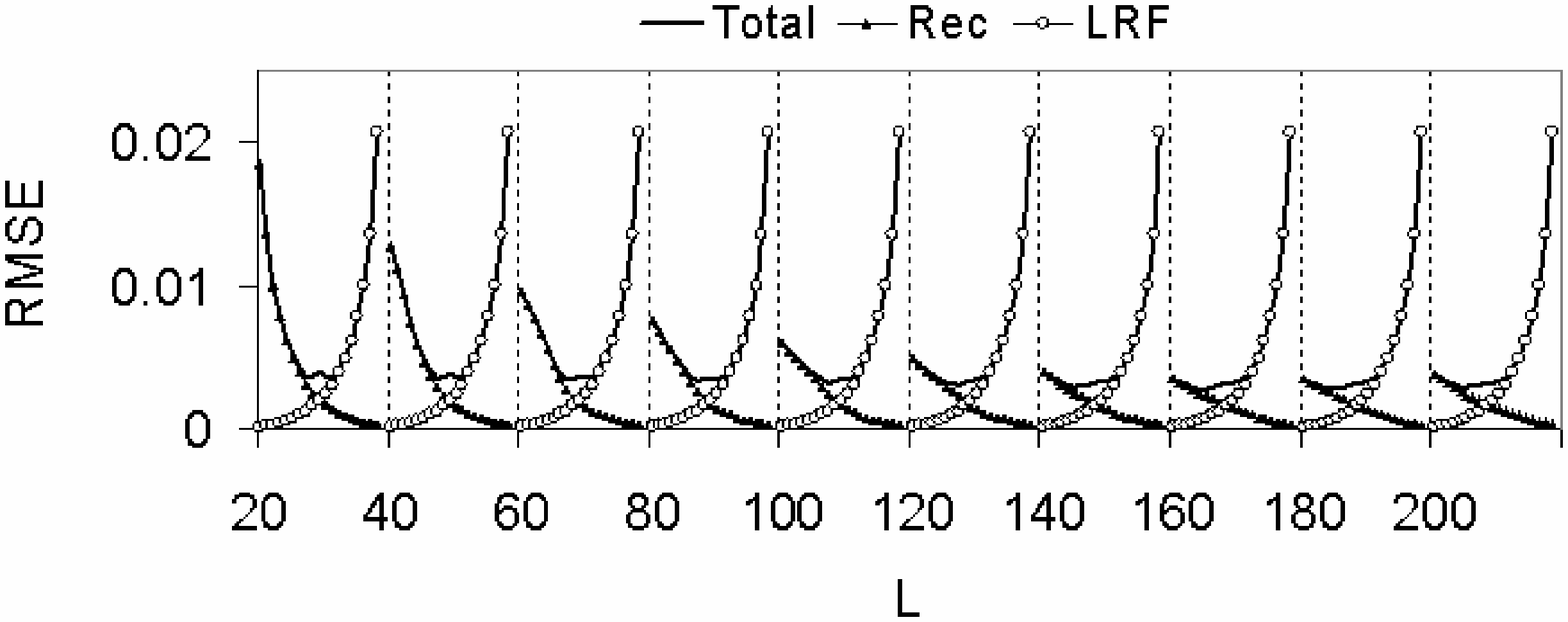}
                    \vspace{-3mm}\caption[t]{RMSE of forecast: t.s.~\eqref{eq:white} with $b<1$,\\ different $L_{\mathrm{rec}}$}
                    \label{fig:for_exp1_cos_noise}
             \end{figure}
             \begin{figure}[h!]
                    \includegraphics[width = 70mm]{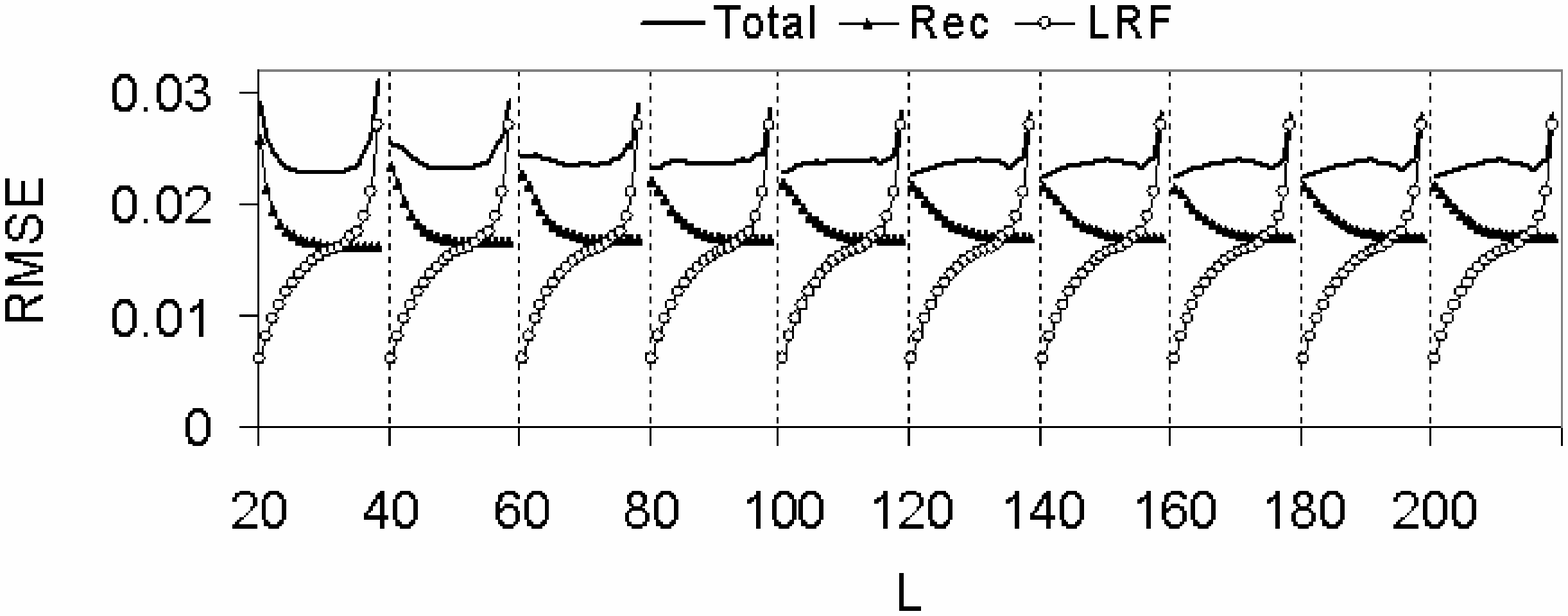}
                    \vspace{-3mm}\caption[t]{RMSE of forecast: t.s.~\eqref{eq:white} with $b>1$,\\ different $L_{\mathrm{rec}}$}
                    \label{fig:for_exp2_cos_noise}
             \end{figure}
\end{center}

The deterministic residual can provide a specific behavior of errors.
Specifically, for the time series~\eqref{eq:1-1} we obtain that the choice of an even $N$ and odd values of $L_{\mathrm{LRF}}$
provides a decrease of the  errors of projectors as the window length increases (see Fig.~\ref{fig:sawP}).
Therefore, for such choice of $L$ and $N$ we observe similar behavior for two sources of forecast errors
and the optimal choice is that the window length should be as large as possible.

\section{Subspace-based methods of parameter estimation}
\label{sec:estimation}

While the problems of reconstruction and forecasting are traditionally included into
the scope of problems solved by SSA, estimation of signal parameters by the subspace-based methods
 is often not considered within the framework of SSA. Therefore, we will describe the subspace-based methods
in more detail to demonstrate their cohesion with SSA.

\subsection{Basic facts}
\label{subsec:estimation:basic}

\begin{definition}
\label{l2}
The companion matrix of a polynomial $p(\mu)=\mu^n+c_1\mu^{n-1}+\ldots+c_{n-1}\mu+c_n$ is
\begin{eqnarray*}
\mathbf{C}=
\left(
\begin{array}{lllll}
 0&0&0&\ldots&-c_n\cr
 1&0&0&\ldots&-c_{n-1}\cr
 0&1&\ldots&0&-c_{n-2}\cr
 \vdots&\vdots&\ddots&\vdots&\vdots\cr
 0&0&\ldots&1&-c_1\cr
\end{array}
\right).
\end{eqnarray*}
\end{definition}

\begin{proposition}
\label{prop:roots}
The roots of a polynomial coincide with the eigenvalues of its companion matrix.
\end{proposition}

%\br
Note that the multiplicities of roots are equal to the algebraic multiplicities of the eigenvalues
of the companion matrix (i.e., to the multiplicities of roots of the matrix characteristic polynomial).
However, these multiplicities do not always coincide with the geometric multiplicities
equal to the dimensions of the eigenspaces corresponding to the eigenvalues.
%\er

To derive the analytic form \eqref{eq:GEN_REQ} of the signal we need to find roots of the characteristic polynomial of
the LRF which governs the signal.
By Proposition~\ref{prop:roots}, we have to  find either the roots of the characteristic polynomial
or the eigenvalues of its companion matrix. The latter does not require a linear recurrent formula
itself. Let us demonstrate that to find the signal roots it is sufficient to know the basis of the signal trajectory space.

Let $\mathbf{C}$ be a full-rank $d\times d$ matrix, $X\in \mathbb{R}^d$, and $\mathbf{X}$ be a full-rank $L\times d$ matrix, $L > d$,
which can be expressed as
\begin{equation}
\label{l3}
 \mathbf{X}=
 \left(
 \begin{array}{l}
 X^{\rm T}\cr
 X^{\rm T} \mathbf{C}\cr
 %X^{\rm T} \mathbf{C}^2\cr
 \vdots \cr
 X^{\rm T} \mathbf{C}^{L-1}
 \end{array}\right).
\end{equation}

Let us again denote the matrix $\mathbf{X}$ without the last row by $\last{\mathbf{X}}$  and  the
matrix $\mathbf{X}$ without its first row by $\first{\mathbf{X}}$.
It is clear that $\first{\mathbf{X}}=\last{\mathbf{X}}\mathbf{C}$. We call this property of $\mathbf{X}$ \emph{shift property}
given by the matrix $\mathbf{C}$.

\begin{proposition}
\label{eq:shift}
Let $\mathbf{X}$ satisfy the shift property given by the matrix $\mathbf{C}$,
$\mathbf{P}$ be a full-rank $d\times d$ matrix, and $\mathbf{Y}=\mathbf{X}\mathbf{P}$.
Then the matrix $\mathbf{Y}$ satisfies the shift property
given by the matrix $\mathbf{D}=\mathbf{P}^{-1}\mathbf{C}\mathbf{P}$, i.e.,
$\first{\mathbf{Y}}=\last{\mathbf{Y}}\mathbf{D}$.
\end{proposition}

\textbf{{Proof.}}
\begin{eqnarray*}
\mathbf{X}\mathbf{P}&=&
 \left(
 \begin{array}{l}
 X^{\rm T}\mathbf{P}\cr
 X^{\rm T}\mathbf{C}\mathbf{P}\cr
 %X^{\rm T}\mathbf{C}^2\mathbf{P}\cr
 \vdots\cr
 X^{\rm T}\mathbf{C}^{L-1}\mathbf{P}\cr
 \end{array}
 \right)=
 \left(
 \begin{array}{l}
 X^{\rm T}\mathbf{P}\cr
 X^{\rm T}\mathbf{P}(\mathbf{P}^{-1}\mathbf{C}\mathbf{P})\cr
 %X^{\rm T}\mathbf{P}(\mathbf{P}^{-1}\mathbf{C}^2\mathbf{P})\cr
 \vdots\cr
 X^{\rm T}\mathbf{P}(\mathbf{P}^{-1}\mathbf{C}^{L-1}\mathbf{P})\cr
 \end{array}
 \right)\\
&=& \left(
 \begin{array}{l}
 X^{\rm T}\mathbf{P}\cr
 X^{\rm T}\mathbf{P}\mathbf{D}\cr
 %X^{\rm T}\mathbf{P}\mathbf{D}^2\cr
 \vdots\cr
 X^{\rm T}\mathbf{P}\mathbf{D}^{L-1}\cr
 \end{array}
 \right)=
 \left(
 \begin{array}{l}
 Y^{\rm T}\cr
 Y^{\rm T}\mathbf{D}\cr
 %Y^{\rm T}\mathbf{D}^2\cr
 \vdots\cr
 Y^{\rm T}\mathbf{D}^{L-1}\cr
 \end{array}
 \right),
\end{eqnarray*}
where $Y=X^{\rm T}\mathbf{P}$.
This implies the shift property for $\mathbf{Y}$.
\hfill $\blacksquare$

%\br
Note that multiplication by a nonsingular matrix $\mathbf{P}$ can be considered as a change of coordinates
in the column space of the matrix  $\mathbf{X}$.
%\er

%\br
It is easily seen that the matrices $\mathbf{C}$ and $\mathbf{D}=\mathbf{P}^{-1}\mathbf{C}\mathbf{P}$ have the same eigenvalues;
these matrices are called \emph{similar}.
%\er

%\begin{remark}
\br
%\label{eq:shift}
Let the matrix $\mathbf{Y}$ satisfy the shift property given by the matrix $\mathbf{D}$.
Then $\mathbf{D}=\last{\mathbf{Y}}^{\dagger}\first{\mathbf{Y}}$, where ${\mathbf{A}}^{\dagger}$ denotes the Moore-Penrose pseudoinverse
of ${\mathbf{A}}$.
\er
%\end{remark}

\begin{proposition}
\label{prop:shift_basis}
Let a time series $F$ satisfy the minimal LRF  \eqref{eq:lrf} of order $d$,  $L \ge d$ be the window length,
$\mathbf{C}$ be the companion matrix of the characteristic polynomial of this LRF.
Then any $L\times d$ matrix $\mathbf{Y}$ with columns forming
a basis of the trajectory space of $F$
satisfies the shift property given by some matrix $\mathbf{D}$.
Moreover, the eigenvalues of this shift matrix $\mathbf{D}$ coincide with the eigenvalues of the companion matrix $\mathbf{C}$ and, therefore,
with the roots of the characteristic polynomial of the LRF.
\end{proposition}

\textbf{Proof.}
Note that for any $0\leq i <N-1$ we have
\begin{eqnarray*}
(f_i, f_{i+1}, \ldots, f_{i+(d-1)}) \mathbf{C} = (f_{i+1}, f_{i+2}, \ldots , f_{i+d}).
\end{eqnarray*}
Therefore, \eqref{l3} holds for $X=(f_0, f_{1}, \ldots,  f_{d-1})^\mathrm{T}$.
It is known that for a time series governed by a minimal LRF of order $d$, any $d$ adjacent vectors of the embedding are independent.
Consequently, the matrix $\mathbf{X}$ is of full rank and we can apply Proposition~\ref{eq:shift}.
\hfill $\blacksquare$

%\begin{remark}
\br
The SVD of the $L$-trajectory matrix of a time series provides a basis of
its trajectory space. Namely, the left singular vectors which correspond
to the nonzero singular values form such a basis.
If we observe a time series `signal + residual', then the SVD of its $L$-trajectory matrix
provides the basis of the signal subspace
under the condition of exact strong separability of the signal and the residual, see \cite{Golyandina.etal2001}.
\er
%\end{remark}

\subsection{ESPRIT}
\label{subsec:estimation:esprit}
Consider a time series $F_N=\{f_i\}_{i=0}^{N-1}$, $f_i=s_i+r_i$, where $S_N=\{s_i\}_{i=0}^{N-1}$ is a time series governed
by a linear recurrent formula of order $r$ (i.e. signal), $R_N=\{r_i\}_{i=0}^{N-1}$ is a residual (noise, perturbation).
Let again $\mathbf{X}$ be the trajectory matrix of $F_N$.
In the case of exact or approximate separability of the signal and the residual and if the signal is dominant,
the subspace $\sspan\{U_1,\ldots, U_r\}$ can be considered as an estimate of the true signal subspace.
Therefore, we can use $\widetilde{\mathbf{Y}}=\mathbf{U}_r=[U_1:\ldots :U_r]$ as an estimate of $\mathbf{Y}$
 from Proposition~\ref{prop:shift_basis}. Then the shift property is fulfilled
only approximately and $\last{\mathbf{U}_r}\mathbf{D}\approx \first{\mathbf{U}_r}$.

Let us study the methods of finding the matrix $\mathbf{D}$.
The idea was introduced in \cite{Kung.etal1983} devoted to the problem of estimating the frequencies in a sum of sinusoids,
in the presence of noise.
The method was given its well-known name ESPRIT in \cite{Roy.Kailath1989}, which was later used in many other papers devoted to
the DOA (Direction of Arrival) problem. This method is also called Hankel SVD (HSVD \cite{Barkhuijsen.etal1987}, for the LS version)
and Hankel Total Least Squares (HTLS \cite{VanHuffel.etal1994}, for the TLS version).

\subsubsection{Least Squares (LS-ESPRIT), HSVD}
The LS-ESPRIT estimate of the matrix ${\mathbf{D}}$ is
\begin{equation}
\widehat{\mathbf{D}}=\last{\mathbf{U}_r}^{\dagger}\first{\mathbf{U}_r}=
(\last{\mathbf{U}_r}^{\mathrm{T}}\last{\mathbf{U}_r})^{-1}\last{\mathbf{U}_r}^{\mathrm{T}}\first{\mathbf{U}_r}.
\end{equation}

%\begin{remark}
The eigenvalues of $\widehat{\mathbf{D}}$ do not depend on the choice of the basis of the subspace $\sspan\{U_1,\ldots, U_r\}$.
In fact, if we change the coordinates so that $\mathbf{W}=\mathbf{U}_r\mathbf{P}$, where $\mathbf{P}$ is a nondegenerate $r\times r$ transfer matrix,
then $\last{\mathbf{W}}=\last{\mathbf{U}_r}\mathbf{P}$ and $\first{\mathbf{W}}=\first{\mathbf{U}_r}\mathbf{P}$.
Hence, by the direct substitution we have that the estimators of ${\mathbf{D}}$ obtained by using $\mathbf{W}$ or $\mathbf{U}_r$
are similar matrices and therefore have the same eigenvalues.
%\end{remark}

\subsubsection{Total Least Squares (TLS-ESPRIT), HTLS}
Since $\mathbf{U}_r$ is known only approximately, then
there are errors in both $\last{\mathbf{U}_r}$ and $\first{\mathbf{U}_r}$.
Therefore, the solution of the approximate equality $\last{\mathbf{U}_r}\mathbf{D}\approx \first{\mathbf{U}_r}$ based on
the Total Least Squares method can be more accurate.

Let us recall that to solve the equation $\mathbf{A}\mathbf{X}\thickapprox\mathbf{B}$ TLS minimizes the following sum:
\begin{eqnarray}
\label{eq:TLSproblem}
\|\widetilde{\mathbf{A}}-\mathbf{A}\|^2_\mathrm{F}+\|\widetilde{\mathbf{B}}-\mathbf{B}\|^2_\mathrm{F}
\longrightarrow \min,\ \mbox{where}\\
\nonumber
(\widetilde{\mathbf{A}},\widetilde{\mathbf{B}})\in \{(\widetilde{\mathbf{A}},\widetilde{\mathbf{B}}):
\exists \mathbf{Z}, \widetilde{\mathbf{A}}\mathbf{Z}=\widetilde{\mathbf{B}}\}.
\end{eqnarray}

Set $\mathbf{A}=\last{\mathbf{U}_r}$, $\mathbf{B}=\first{\mathbf{U}_r}$ in \eqref{eq:TLSproblem}.
Then the matrix $\mathbf{Z}$ that minimizes \eqref{eq:TLSproblem} is called the TLS-estimate of $\mathbf{D}$
(see \cite{Groen1996} for explicit formulas).

Let us consider the dependence of the TLS-ESPRIT solution on the choice of the basis
of $\sspan\{U_1,\ldots ,U_r\}$.
Generally speaking (computer experiments confirm this), this dependence takes place.
However, the following assertion can be proved.

\begin{proposition}
The TLS-ESPRIT solution does not depend on the unitary transformation of the basis of $\sspan\{U_1\ldots U_r\}$.
In particular, the TLS-ESPRIT estimate is the same for any orthonormal basis.
\end{proposition}

Indeed, suppose that the minimum in \eqref{eq:TLSproblem} is achieved at some $\mathbf{Z}$.
Let us consider the revised problem \eqref{eq:TLSproblem} with replacement of $\mathbf{A}$ and $\mathbf{B}$ by
$\mathbf{A}\mathbf{P}$ and $\mathbf{B}\mathbf{P}$, respectively.
It is easy to see that if $\mathbf{P}$ is an $r\times r$ unitary matrix (and therefore preserves norms),
then the solution of the revised problem \eqref{eq:TLSproblem} has the form of $\mathbf{P}^{-1}\mathbf{Z}\mathbf{P}$, which is similar to $\mathbf{Z}$.
Thus, the eigenvalues of the solution of the problem \eqref{eq:TLSproblem} do not depend on simultaneous rotations/reflections
in the column spaces of the matrices $\mathbf{A}$ and $\mathbf{B}$.

\subsubsection{DOA and time series analysis}

Recall that ESPRIT-like methods can be applied both  to the general problem of
parameter estimation for the time series and to the DOA problem specifically. However, there are
several special aspects of application to DOA:
\begin{enumerate}
\item Input data is in the form of an $L\times K$ matrix (an analogue to the trajectory matrix). The matrix satisfies the shift property but
it is not necessarily a Hankel matrix.
\item The number of sensors ($L$) is fixed and is usually not large. Therefore, the properties of the method are considered
under the condition that $L$ is fixed and $K\rightarrow \infty$.
\item The problem can be stated so that
the entries of the input matrix are noisy with independent noise realizations
(this does not hold for the trajectory matrices where both the signal and noise form Hankel matrices).
\item The data for DOA is mostly complex-valued with
complex circular white Gaussian noise.
%(Definition: a random variable $x$ is called circular if $\mathbb{E} x=0$, $\mathbb{E} x x = 0$, $\mathbb{E} x x^*=\sigma^2$.)
\end{enumerate}

Generally, these aspects can influence the statement of the parameter choice problem and the rule of the optimal parameter choice.

\subsubsection{Dependence on the window length}
The paper \cite{Badeau.etal2008} contains the following theoretical result: while estimating the frequency of
a noisy sinusoid, the asymptotic ($N\rightarrow \infty$) variance of the first-order error has order $1/(K^2 L)$ and is symmetric
with respect to $N/2$.
Therefore, the asymptotic optimal window length is equal to $N/3$ or $2N/3$.
Numerical experiments confirm this conclusion. Let us remark that it is not always true that the first-order
(with respect to the perturbation level) error
is the main-term error as the time series length tends to infinity.
Therefore, it is better to check the correspondence between the first-order error and the total error through simulation.

In \cite{Djermoune.Tomczak2009} an explicit form of the asymptotic variance of the first-order error
is derived in the general case of damped complex exponentials.  In the case of undamped complex exponentials,
the derived form coincides with that in \cite{Badeau.etal2008}. As for damped complex exponentials, the result is
that the optimal window length lies between $N/3$ and $N/2$ and approaches $N/2$ as the damping factor increases.
It is shown in \cite{Djermoune.Tomczak2009} that for $s_n=\exp((\alpha+i\beta)n)$, $i=\sqrt{-1}$, the first-order variances of the ESPRIT estimates
of $\alpha$ and $\beta$ are equal.
Therefore, the optimal window lengths are the same for estimators of the damping factor $\alpha$ and of the frequency $\beta$ .

In the previous sections we demonstrated that the separability of the signal from deterministic and stochastic residuals has a different
nature and therefore leads to different consequences. Let us consider how this difference reveals itself in the problem of the frequency estimation
by ESPRIT. We perform simulations for the time series \eqref{eq:det}--\eqref{eq:red} with $c=\sigma=0.1$, $b=1$, $N=100$.

\begin{center}
             \begin{figure}[h!]
                    \includegraphics[width = 5.8cm]{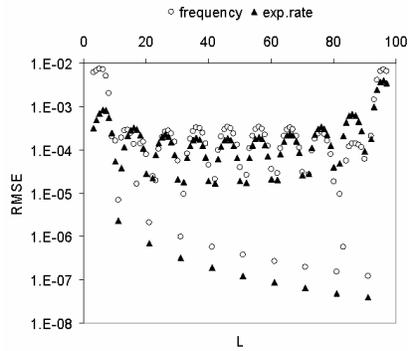}
                    \vspace{-3mm}\caption[t]{RMSE of frequency and exp.rate estimates: t.s.~\eqref{eq:det} (log-scale)}
                    \label{fig:cos_constFRlog}
             \end{figure}
\end{center}

\begin{center}
             \begin{figure}[h!]
                    \includegraphics[width = 5.8cm]{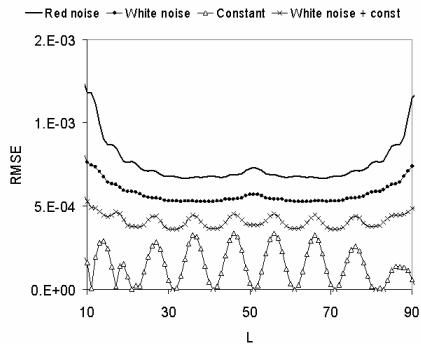}
                    \vspace{-3mm}\caption[t]{RMSE of frequency estimates: different types of residuals, t.s.~\eqref{eq:det}--\eqref{eq:red}, $L\sim N/2$}
                    \label{fig:cf_res_MSE}
             \end{figure}
\end{center}

Fig.~\ref{fig:cos_constFRlog} contains the results for the deterministic perturbation, including
the specific behavior of RMSE in the case of one-sided (left) orthogonality.
Fig.~\ref{fig:cf_res_MSE} contains RMSE of frequency estimates for different kinds of residuals.
The behavior of errors is very similar to that in the signal reconstruction, see Fig.~\ref{fig:cf_resMSE}.
Also, the errors of frequency and exponential rate (damping factor) estimates are approximately equal if $L$ is proportional to $N$.
The main difference from the reconstruction errors is in the size of errors, which is much smaller.
Therefore we use 1000 instead of 100 simulated series to estimate RMSE with sufficient accuracy.

\begin{center}
             \begin{figure}[h!]
                    \includegraphics[width = 5.8cm]{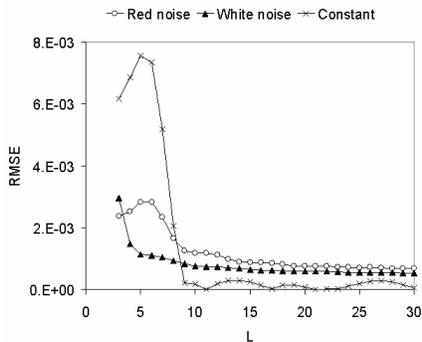}
                   \vspace{-3mm}\caption[t]{RMSE of frequency estimates: different types of residuals, t.s.~\eqref{eq:det}--\eqref{eq:red}, $L$ is small}
                    \label{fig:cf_res_MSE_smallL}
             \end{figure}
\end{center}

Fig.~\ref{fig:cf_res_MSE_smallL} focuses attention on the error behavior for small window lengths.
One can see that the dominance of errors for the time series \eqref{eq:det} with deterministic residuals
is a distinctive feature.

\vspace{-10mm}

\vspace{-10mm}

\subsection{Brief review of other subspace-based methods}
\label{subsec:estimation:other}
In this subsection, we briefly describe several subspace-based methods in addition to ESPRIT-like ones.
These methods are applied to time series governed by LRFs and in fact estimate the main (signal)
roots of the corresponding characteristic polynomials.
The basic subspace-based methods were developed for the case of a noisy sum of imaginary exponentials (cisoids)
or of real sinusoids, for the purpose of frequency estimation, see e.g. \cite{Stoica.Moses1997}.
We are mostly interested in the methods that can be applied to any time series
of finite rank given in the form \eqref{eq:GEN_REQ}.

We start with the description of general methods in the complex-valued case.

\smallskip\noindent
\textbf{Version 1} Consider an LRF that governs the signal (the best choice is the min-norm LRF, see Section \ref{subsec:LRF},
however this is not essential). Then we find all roots $\mu_m$ of the characteristic polynomial of this LRF and then find coefficients $c_{mj}$
in \eqref{eq:GEN_REQ}.
The coefficients $c_{mj}$ corresponding to the extraneous roots are equal to 0. In the case of a noisy signal, $\widehat{\mu}_m$ are the roots of a polynomial
with coefficients from a vector that belongs to $\mathcal{L}_r^\bot$, and the extraneous roots have small absolute values
of the LS estimates $\widehat{c}_{mj}$.

\smallskip\noindent
\textbf{Version 2} Let us consider the forward and backward min-norm predictions. It is known that
the cor\-res\-pond\-ing characteristic polynomials have the conjugate extraneous roots and their signal roots
are connected by the relation $z'=z^*/\|z\|^2$. Note that the forward prediction given
by a vector $A\in\mathcal{L}_r^\bot$ corresponds to the roots of
$\langle Z(z), A \rangle =0$, where $Z(z)=(1,z,\ldots,z^{L-1})^\mathrm{T}$ and $\langle \cdot\,  ,\cdot \rangle$ is the inner product in $\mathbb{C}$.
At the same time,  the backward prediction
given by a vector $B\in\mathcal{L}_r^\bot$ corresponds to the roots of $\langle Z(1/z),B \rangle=0$.
If we consider the roots of the forward and backward min-norm polynomials together, then all the extraneous roots lie inside the unit circle,
while one of $z'$ and $z$ is located on or beyond the unit circle. This allows us to detect the signal roots.

\smallskip\noindent
\textbf{Version 3} Let us take a set of vectors from $\mathcal{L}_r^\bot$.
Each vector from $\mathcal{L}_r^\bot$ with a nonzero last coordinate generates
an LRF. The signal roots of the characteristic polynomials of these LRFs are equal (or are close for
a noisy signal), whereas the extraneous roots are arbitrary. Therefore, the signal roots correspond
to clusters of roots if we consider pooled roots. One of the ways to choose vectors from
$\mathcal{L}_r^\bot$ is to take the set of eigenvectors corresponding to noise.

\medskip
There are some other methods that are developed for estimating frequencies in a noisy sum of undamped sinusoids or complex exponentials.
Let for simplicity $s_n=\sum_{k=1}^r c_k e^{2\pi i \omega_k n}$.
In this case, the signal roots $e^{2\pi i \omega_k}$ have absolute values equal to 1 and can be parameterized only by one parameter (frequency).
Let $W=W(\omega)=Z(e^{2\pi i \omega})$. Since
$W(\omega_k)\in \mathcal{L}_r^{(\mathrm{s})}$, $\langle W(\omega_k), A \rangle =0$ for all $A\in \mathcal{L}_{r+1,L}^{(\mathrm{s})}$.
If $A\in \mathcal{L}_r^\bot$, then we can consider
the square of the cosine  of the angle between $W(\omega_k)$ and $A$ as a measure of their orthogonality. This idea forms
the basis for the Min-Norm and MUSIC methods.
The names of methods in which roots are ordered by the absolute value of the deviation of their modules from the unit circle
begin with `root-'.
%In fact, only in the `root-' methods it is necessary to calculate roots of polynomials.

\smallskip\noindent
\textbf{Version 4. Min-Norm} Let $f(\omega)=\cos^2(\widehat{W(\omega),A})$, where $A= \mathrm{P}_r^\bot \mathbf{e}_L \in \mathcal{L}_r^\bot$ is
the vector corresponding to the min-norm forward prediction. The Min-Norm method consists in searching for the maximums of $1/f(\omega)$,
and the function $1/f(\omega)$ of $\omega$ is interpreted as a so-called pseudospectrum with peaks at the frequencies presented in the signal.

\smallskip\noindent
\textbf{Version 5. Root Min-Norm} We consider the min-norm LRF and choose the $r$ closest to
the unit circle roots of its characteristic polynomial.

\smallskip\noindent
\textbf{Version 6. MUSIC}  Let $f(\omega)=\cos^2(\widehat{W(\omega),\mathcal{L}_r^\bot})$. If we take eigenvectors
$U_j$, $j=r+1,\ldots,L$, as a basis of $\mathcal{L}_r^\bot$, then $\mathbf{U}_{r+1,L} \mathbf{U}^*_{r+1,L}$
provides the matrix of projection on $\mathcal{L}_r^\bot$ and therefore
$f(\omega)=W^*(\mathbf{U}_{r+1,L} \mathbf{U}^*_{r+1,L}) W/\|W\|^2=\suml_{j=r+1}^L f_j(\omega)$,
where $f_j(\omega)=\cos^2(\widehat{W(\omega),U_j})$.
Thus, the MUSIC method can be considered from the viewpoint of the subspace properties and does not require the computation
of roots of characteristic polynomials.
Similar to  the Min-Norm method, the MUSIC method  consists in searching for the maximums of the pseudospectrum $1/f(\omega)$.

There is a modification of MUSIC called `EV'. In this modification, the pseudospectrum is constructed on the base of
the weighted sum $\sum_{j=r+1}^L f_j(\omega)/\lambda_j$. The EV method can improve frequency estimates if
the signal rank $r$ is detected with error, since the weights decrease the contribution of
summands corresponding to the noise eigenvectors that are adjacent to the signal ones.
Note that the EV method is not expressed in terms of the signal subspace.

\smallskip\noindent
\textbf{Version 7. Root-MUSIC} This method involves calculation of the polynomial roots.
Specifically, we calculate the roots of a polynomial of $z$ solving the equation
$(Z(1/z))^\mathrm{T}\mathbf{U}_{r+1,L} \mathbf{U}_{r+1,L}^* Z(z)=0$.
This polynomial can be considered as the multiplication of polynomials with signal roots connected by the relation
$z'=z^*/\|z\|^2$, that is, the forward and backward predictions.
Then the roots with modules less than or equal to 1 are assembled in the ascending order of their modules.
The signal roots have modules that are the closest to 1.

\section{The rate of convergence}
\label{sec:convergence}

The paper \cite{Nekrutkin2010} contains theoretical results on convergence (as the time series length $N$ tends to infinity) for the methods that
are based on the estimation of the signal subspace. Here we investigate the rate of convergence by means of examples.

Let us consider two time series with lengths $N_1$ and $N_2$ such that $N_2= 4 N_1$.
Let RMSE be the measure of accuracy. If the residual $R_N$ is random, then we perform simulations to estimate
RMSE. We denote the ratio of RMSEs for the window lengths $N_1$ and $N_2$ by $\Delta=\mathrm{RMSE}_1/\mathrm{RMSE}_2$.
Then, $\Delta=8$ indicates the rate of convergence $1/N^{1.5}$, $\Delta=2$ corresponds to the rate of convergence $1/N^{0.5}$,
and $\Delta=1$ means that there is no convergence at all.
To estimate $\Delta$, we use $N_1=6399$ and $N_2=25599$ (we chose odd time series lengths to
consider $(N+1)/2$ as one of window lengths)
%obtain the symmetry for the window lengths about $(N+1)/2$).

We discuss examples with the following three kinds of perturbation of the signal:
by a constant,  by noise and by a sum of noise and a constant.
Also, we consider two types of random noise, white and red.
We perform numerical experiments for the time series \eqref{eq:det}--\eqref{eq:red} with $\sigma=0.1$,
$b=1$, $\alpha=0.5$.
In what follows we calculate the frequency and exponential base estimates using the LS-ESPRIT method
(the difference with the results of TLS-ESPRIT is small and does not influence the conclusions).

Tables~\ref{tab:rec}--\ref{tab:freq} include the results on convergence based on 1000 simulations. The column `c' corresponds to
the time series~\eqref{eq:det} with constant residuals, the columns `wn' and `rn' contain the results for white-noise (the time series~\eqref{eq:white})
and red-noise (the time series~\eqref{eq:red}) residuals respectively, and the column `c+wn' includes estimates of $\Delta$ for the time series
\eqref{eq:mix} with combined perturbation.

\begin{table}
             \begin{minipage}{70mm}
\caption{The rate of convergence of reconstruction: estimated $\Delta$}
\label{tab:rec}
\begin{tabular}{|c||c|c|c|c|}
  \hline
  $L$         & c   & c+wn& wn & rn   \\
  \hline
  %\hline
  $r+1$       &\noc{1.0} & \noc{1.0}&\noc{1.0} & \noc{1.0}\\
  $20$        &--- & \noc{1.0}&\noc{1.0} & \noc{1.0}\\
  $25$        &\noc{1.0} & \noc{1.0}&\noc{1.0} & \noc{1.0}\\
  \hline
  $(N+1)/2-5$ &8.0 & 1.9&2.0 & 2.1\\
  $(N+1)/2$   &--- & 1.9&2.0 & 2.1\\
  \hline
\end{tabular}
             \end{minipage}
             \hfil
             \begin{minipage}{70mm}
 \caption{The rate of convergence of projector estimates: estimated $\Delta$}
\label{tab:proj}
\begin{tabular}{|c||c|c|c|c|}
  \hline
  $L$         & c   & c+wn& wn & rn   \\
  \hline
  $r+1$       & \noc{1.0} & \noc{1.0} & 2.2 & \noc{1.0} \\
  $20$        & --- & 3.1 & 2.8 & \noc{1.1} \\
  $25$        & \noc{1.0} & \noc{1.0} & 3.0 & \noc{1.1} \\
  \hline
  $(N+1)/2-5$ & 4.0 & 2.0 & 2.0 & 2.0 \\
  $(N+1)/2$   & --- & 2.0 & 2.0 & 2.0 \\
  \hline
\end{tabular}
             \end{minipage}
\end{table}

\begin{table}
             \begin{minipage}{70mm}
\caption{The rate of convergence of expon. base estimates: estimated $\Delta$}
\label{tab:base}
\begin{tabular}{|c||c|c|c|c|}
  \hline
  $L$         & c   & c+wn& wn & rn   \\
  \hline
  %\hline
  $r+1$       & --- & 4.1&4.1 & 4.0\\
  $20$        & --- & 4.0&4.1 & 3.9\\
  $25$        & \noc{1.0} & 2.8&4.1 & 3.9\\
  \hline
  $(N+1)/2-5$ & 16.0& 8.2&8.2 & 8.1\\
  $(N+1)/2$   & --- & 8.2&8.2 & 8.1\\
  \hline
\end{tabular}
             \end{minipage}
             \hfil
             \begin{minipage}{70mm}
\caption{The rate of convergence of frequency estimates: estimated $\Delta$}
\label{tab:freq}
\begin{tabular}{|c||c|c|c|c|}
  \hline
  $L$         & c   & c+wn& wn & rn   \\
  \hline
  %\hline
  $r+1$       & \noc{1.0} & \noc{1.0}&2.0 & \noc{1.0}\\
  $20$        & --- & 2.7&2.5 & \noc{1.0}\\
  $25$        & \noc{1.0} & \noc{1.0}&2.7 & \noc{1.0}\\
  \hline
  $(N+1)/2-5$ &16.0 & 8.2&8.4 & 8.2\\
  $(N+1)/2$   & --- & 8.2&8.4 & 8.2\\
  \hline
\end{tabular}
             \end{minipage}
\end{table}

It would appear reasonable that  the convergence rates for fixed window lengths and for window lengths proportional to $N$
differ. Also, the multiplicity of window lengths to the period of the sine-wave signal (10 in the considered examples) can be important.
Therefore, we analyze two sets of window lengths. The first set includes fixed window lengths:
the minimal $L=r+1$, where $r=2$ is the rank of sinusoid, $L=20$ is divisible by 10, and $L=25$ is a common case.
The second set contains two window lengths close to $N/2$: $L=(N+1)/2$ is divisible by 10 and $L=(N+1)/2-5$ is a common case.
Note that if there is the exact separability (that is possible in the case of deterministic residuals only), then the ratio $\Delta$ cannot
be calculated (the sign `---' in the tables). We do not consider the windows length $L>(N+1)/2$, since these values of $L$ lead to either
the same or worse convergence rates in comparison with the window length $N-L+1<(N+1)/2$.

Let us discuss the results presented in Tables~\ref{tab:rec}--\ref{tab:freq}
for window lengths tending to infinity and for fixed window lengths separately.

\medskip\noindent
\textbf{Window length $L\sim N/2$\ }
The simulations provide stable estimates of the convergence rate for the window length $L$ equal to
one-half of the time series length (and more generally, for the window lengths that are proportional
to $N$):

\smallskip
\textbf{(A)} ``signal+noise'' (the time series~\eqref{eq:white}, \eqref{eq:red})\\
(a) the convergence rate of the projector on the signal subspace is $1/N^{0.5}$,\\
(b) the convergence rate of the reconstruction of the whole signal (average error) is $1/N^{0.5}$,\\
(c) the convergence rate of the frequency and exponential rate estimates is $1/N^{1.5}$;

\smallskip
\textbf{(B)} ``signal+constant'' (the time series~\eqref{eq:det})\\
(a) the convergence rate of the projector on the signal subspace is nearly $1/N$,\\
(b) the convergence rate of the reconstruction of the whole signal (average error) is nearly $1/N^{1.5}$,\\
(c) the convergence rate of the frequency and exponential rate estimates is nearly $1/N^2$.

\medskip
Theoretical results on the reconstruction errors in a particular case of a noisy constant signal
\cite{Golyandina.Vlassieva2009}, see \eqref{eq:MSSA}, provide support for a part of the conclusions
derived from the simulations.
Results on RMSE for the  frequency and exponential base estimates are confirmed in \cite{Badeau.etal2008}, where the
case of a noisy sinusoid is considered.
Let us remark that the convergence rate $1/N^{1.5}$ is not surprising, since the Cram\'{e}r-Rao lower bound for
the variance of the frequency estimates has the same order (see, e.g. \cite{Rife.Boorstyn1974, Stoica.etal1997}).
On the other hand, the Cram\'{e}r-Rao lower bound for the variance of estimates of the sinusoid amplitude has the order $1/N^{0.5}$,
which corresponds to the convergence rate of the reconstruction of the signal.
Simulations for the time series \eqref{eq:red} confirm that for $L\sim N/2$ the red-noise residuals provide the same convergence rate
as the white-noise residuals.

One can see that the convergence rate in the examples with pure random residuals is much worse than
that for the example with deterministic (constant) residuals.  As one might expect, the example with combined residuals
inherits the worst case. Simulations confirm
that in the case of a constant residual mixed with a stochastic component (the time series~\eqref{eq:mix}) and the window length $L$ proportional to $N$,
the rate of convergence is the same as for the case \textbf{(A)}.

\medskip\noindent
\textbf{Fixed window length $L=L_0$\ }
Let us consider the case of a fixed window length $L$ and $N\rightarrow \infty$.
The behavior of the rate of convergence is more complicated than the behavior described above.

Analysis of the \emph{reconstruction errors} shows the following behavior:

\smallskip
(\textbf{A}) for ``signal+noise'' (the time series~\eqref{eq:white}, \eqref{eq:red}), there is no convergence to the signal,
even for the  window lengths divisible by the signal (or noise) periods (divisible by 10 in the considered example);

\smallskip
(\textbf{B}) for ``signal+constant'' (the time series~\eqref{eq:det}), the convergence holds only if $L$ (or $K=N-L+1$) is divisible by the period;
in general, there is no convergence.

\medskip
Consequently, if residuals contain noise, there is no convergence.
Thus, in general, small window lengths are not suitable for the problems of signal reconstruction.

\smallskip
Let us now consider the \emph{errors of projector and parameter estimates}.
The difference with the behavior of reconstruction errors consists in the presence of convergence
for small $L$ and white-noise residuals.
However, the simulations for the time series~\eqref{eq:red} with red-noise residuals demonstrate the absence of convergence
for projection and frequency estimation.
%The columns entitled `proj' are related to the errors of estimation of the projector on the signal subspace,
%`freq' and `base' are related to the frequency and exponential base estimates, respectively.
In Tables~\ref{tab:proj} and~\ref{tab:freq}, this absence of convergence
corresponds to the values `1.0' and `1.1' in the column entitled `rn'.
Convergence of the exponential base estimates still takes place, see Table~\ref{tab:base}.

%Since the absence of convergence cannot be worsened,
The question is how the deterministic (constant, in our examples)
compound of the white-noise residuals influences the convergence for small $L$.
Unfortunately, the errors for the projector and parameter estimates converge to 0 for $L=L_0$ only if
the window length is divisible by the signal periods.
%(if both $L$ and $K$ are divisible by the signal periods, then the errors are equal to 0).

Thus, we can conclude that using small window lengths for frequency estimation is possible
only if the residuals are pure white noise, that is, they do not contain deterministic components
and are independent. Otherwise, there is no convergence.

\begin{center}
             \begin{figure}[h!]
                    \includegraphics[width = 6.8cm]{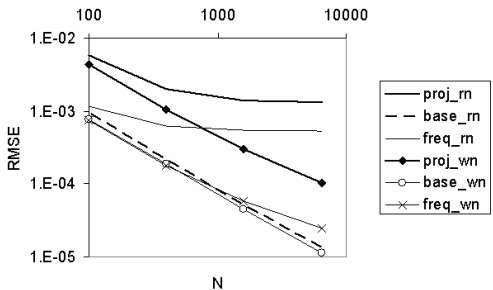}
                    \vspace{-3mm}\caption[t]{Comparison of RMSE for red-noise and white-noise residuals}
                    \label{fig:redwhite_conv}
             \end{figure}
\end{center}

%\begin{remark}
\smallskip
Let us compare the estimation error for white and red noise residuals regardless of the absence of convergence in the latter case.
Fig.~\ref{fig:redwhite_conv} shows the  absolute values of the estimation errors for $L=10$.
The first symbols  of the line titles mean types of the estimated objects: projector (`proj'),
exponential base (`base'), or frequency (`freq'); the last symbols designate types of residuals.
One can see that the sizes of errors for red and white noise residuals are comparable for the considered time series
lengths. This effect is stable enough for different parameters of the time series model.
Thus we can perform estimation with good  accuracy even in the case of the
absence of convergence.
This can be explained by the results of \cite[formula (2.15)]{Nekrutkin2010}, where the upper bounds
for the projector errors are derived. The main term of the upper bound (it is a constant depending on
the sinusoid frequency, the parameters of red noise, and the window length) yields the proper order of errors. Moreover,
\cite[formula (2.9)]{Nekrutkin2010} provides approximately the same
small magnitude $1.3\cdot 10^{-3}$ as presented in Fig.~\ref{fig:redwhite_conv}, the line marked `proj\_rn'.
Calculations confirm  that this magnitude is approximately equal to
$K\big(\mathbf{S}\mathbf{S}^\mathrm{T}\big)^\dag\, \mathbf{\Sigma}\, \big(\mathbf{I}-\mathbf{U}_r^{(s)}(\mathbf{U}_r^{(s)})^\mathrm{T}\big)$,
where $\mathbf{\Sigma}$ is the autocovariance $L\times L$ matrix of the considered red noise. Note that this term
does not converge to 0 as the time series length tends to infinity.
%\end{remark}

\vspace{-5mm}
%\begin{remark}
As we have mentioned, the DOA problems use matrices corresponding to the case of fixed $L=L_0$,
since $L$ is the number of sensors.
Therefore, the estimation of frequencies by ESPRIT-like methods can be performed with high accuracy
only if the residual is pure white noise (or if $L$ is large enough).
 In most existing works devoted to the performance
of the subspace-based methods for the DOA problems, $L$ is fixed (see, for example,
\cite{Krim.etal1992} for the root-MUSIC performance or \cite{Li.etal1993} for a wide class of subspace-based
algorithms including  ESPRIT).
Papers \cite{Badeau.etal2008} and \cite{Djermoune.Tomczak2009} are beyond the scope of DOA
and consider the estimation of signal parameters for an arbitrary noisy signal governed by
an LRF as $\min(L,K)\rightarrow \infty$.
%\end{remark}

\section{Choice of the window length and separability}
\label{sec:noseparability}

\subsection{Modulated sinusoid}

Note that the ability of SSA-like methods to extract exponentially modulated (damped) sinusoids
implies that the SSA method is not simply a spectral method even if we do not apply
SSA for trend extraction.
Note that both damped and undamped sinusoids have SSA-rank 2
(or rank 1 if cisoids are considered in the complex-valued case).
This feature of SSA significantly extends the set of time series that are suitable for the SSA analysis.
Most of the classical methods (e.g. Fourier analysis, seasonal decomposition) deal with either constant amplitudes or with amplitudes proportional
to the trend (if any) and can be reduced to constant amplitudes by transferring time series to the logarithmic scale.
However, time series consisting of several damped sinusoids cannot be reduced to periodic time series
or to multiplicative periodicity. Let us consider, for example, a seasonal component containing both yearly
and quarterly oscillations and let the amplitude of yearly periodicity increase whereas the amplitude
of quarterly oscillations decrease. Then many classical methods fail whereas the SSA-like methods
can easily extract such seasonality.

If the behavior of modulations is more complex than the exponential one, then the SSA-like methods can encounter difficulties.
These methods can still extract such oscillations, however the question of the proper choice of the window length arises.

Let us formulate the question in a more specific form. Consider the signal in the form  of
$s_n=A(n)\cos(2\pi n \omega)$, where $A(n)$ is a slowly (in comparison with $\omega$) varying function.
The question: are there any examples when the choice of the window length
close to $N/2$ is not good.

We consider the time series $f_n=s_n+r_n$ with
\be
\label{eq:mod}
s_n=\cos(2\pi n/19)+ \cos(2\pi n/21),\ r_n=\varepsilon_n.
\ee
Here $s_n=A(n)\cos(2\pi n/20)$ with $A(n)=2\cos(\pi n (1/19-1/21))$ is
a modulated sinusoid of frequency 1/20.
The signal has rank 4 and is asymptotically separable from noise, constant residual and others.
We have two alternative possibilities. The first possibility is to take $L$ close to $N/2$ (e.g., between $N/3$ and
$N/2$) and extract the signal by four leading eigentriples. The second alternative is to take a
window length $L$ so small that the amplitude of the signal is almost constant within the limits of subseries of length
$L$, and then to extract the signal by two leading components.
In the latter case, the left singular vectors are close to the undamped sinusoids and the modulation is caught by
the right singular vectors.

\begin{center}
             \begin{figure}[h!]
                    \includegraphics[width = 72mm]{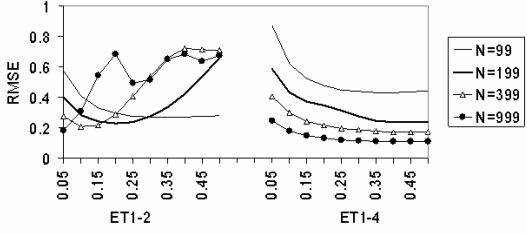}
                    \vspace{-3mm}\caption[t]{RMSE of reconstruction: dependence on $L/N$ for t.s.~\eqref{eq:mod}, ET1--2 and ET1--4}
                    \label{fig:coscos}
             \end{figure}
\end{center}

Numerical simulations (see Fig. \ref{fig:coscos} for the time series lengths
equal to 99, 199, 399, and 999) show that there is no clear choice between the described alternatives.
If the time series length is large enough for approximate separability, then the choice $L\sim N/2$ is better.
Otherwise, window lengths close to a couple of periods provide a better accuracy.
The drawback of the latter choice is that usually we do not know the period and
therefore cannot guess the proper window length.

Thus, under the conditions of approximate separability (one can check if these conditions are met
by means of analysis of the decomposition results, see \cite{Golyandina.etal2001})
the best signal reconstruction uses the window length close to $N/2$ and
the number of the corresponding eigentriples is equal to the signal rank.
The advantages of this choice are the better accuracy and the independence of the window length choice
from the unknown values of periods.
In the considered example, for $L=N/2$, the reconstruction by four eigentriples performs well starting from
$N=160$. However, the choice of $L=N/2$ is appropriate even for $N<120$ if we
produce the reconstruction by two eigentriples. Therefore, there is a limited range of time series lengths
(approximately from 120 to 160), where the choice of the window length close to $N/2$ cannot provide
an adequate accuracy of signal reconstruction.

\vspace{-5mm}
Table~\ref{tab:coscos} contains RMSE for signal reconstruction using two window lengths: $L=40$ (two periods)
and $L=N/2$. The values in bold indicate smaller errors.

\begin{table}[h!]
\caption{RMSE of reconstruction: different parameter choices for t.s.~\eqref{eq:mod}}
\begin{tabular}{|c|c|c|c|}
  \hline
   & $L=40$ & $L=N/2$ & $L=N/2$\\
  N &  2ET &  2ET &  4ET\\
  \hline
  99  &$\mathbf{0.27}$  &$\mathbf{0.27}$   &$0.45$\\
  159 &$\mathbf{0.23}$  &$0.42$   &$\mathbf{0.24}$\\
  199 &$\mathbf{0.22}$  &$0.65$   &$\mathbf{0.25}$\\
  399 &$\mathbf{0.20}$  &$0.70$   &$\mathbf{0.16}$\\
  999 &$0.19$  &$0.68$   &$\mathbf{0.11}$\\
  \hline
\end{tabular}
\label{tab:coscos}
\end{table}

\begin{center}
             \begin{figure}[h!]
             \begin{minipage}{58mm}
                    \includegraphics[width = \linewidth]{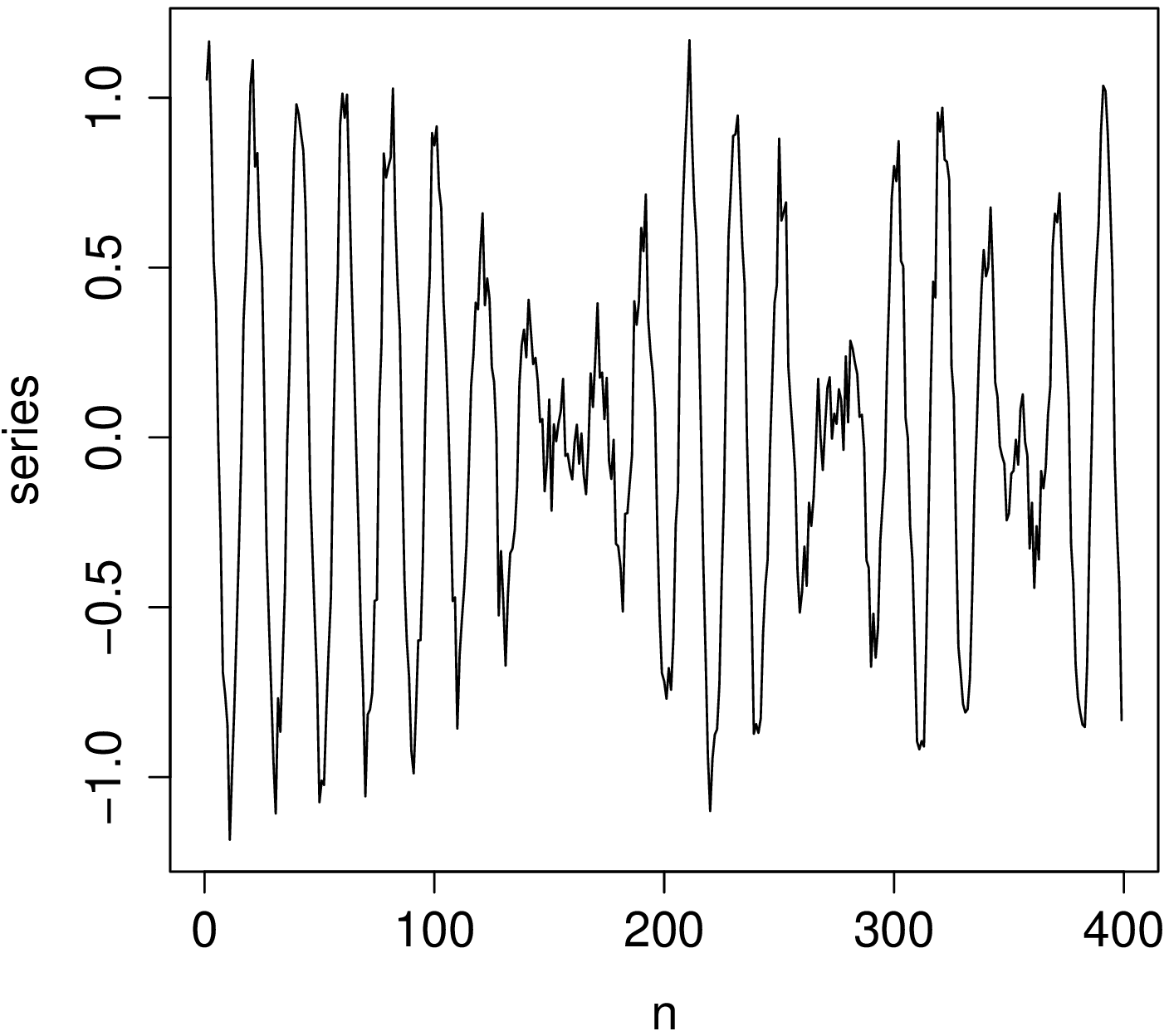}
                    \vspace{-3mm}\caption[t]{Initial time series: t.s.~\eqref{eq:mod2}, $\sigma=0.1$}
                    \label{fig:cos2cos01series}
             \end{minipage}
             \hfil
             \begin{minipage}{58mm}
                    \includegraphics[width = \linewidth]{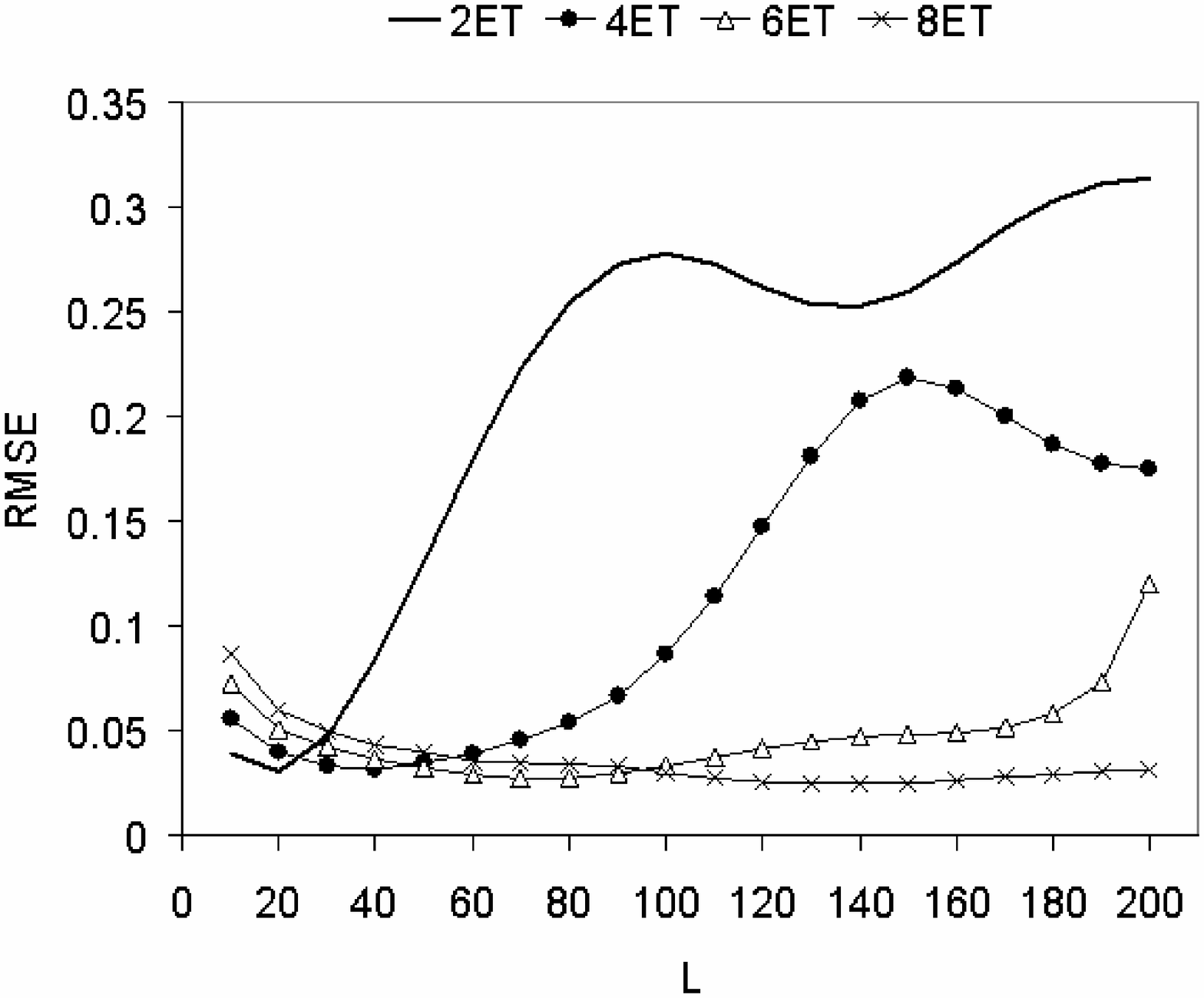}
                    \vspace{-3mm}\caption[t]{RMSE of reconstruction: t.s.~\eqref{eq:mod2}, $\sigma=0.1$}
                    \label{fig:cos2cos01MSE}
             \end{minipage}
             \end{figure}
\end{center}

\begin{center}
             \begin{figure}[h!]
             \begin{minipage}{58mm}
                    \includegraphics[width = \linewidth]{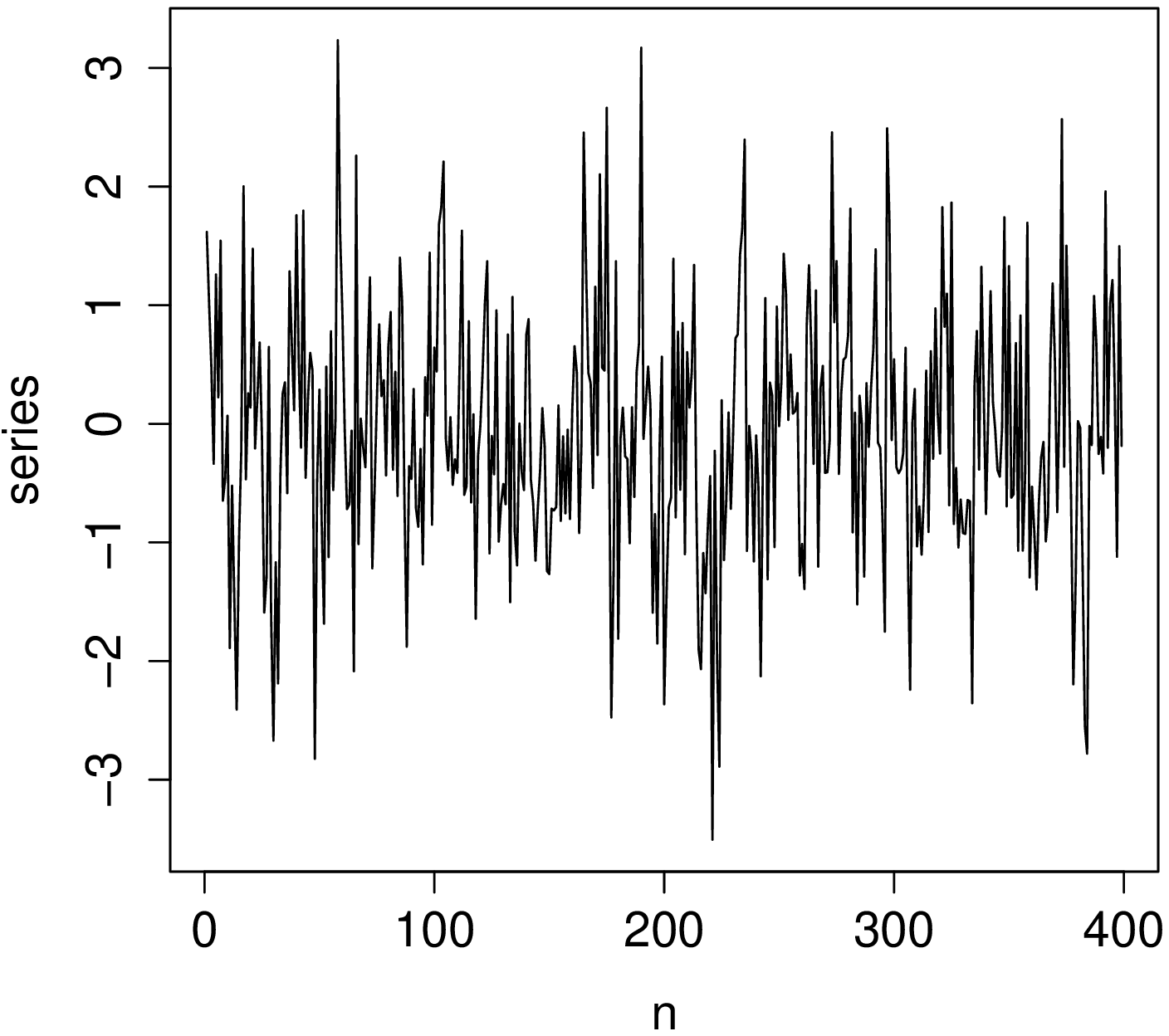}
                    \vspace{-3mm}\caption[t]{Initial time series: t.s.~\eqref{eq:mod2}, $\sigma=1$}
                    \label{fig:cos2cos10series}
             \end{minipage}
             \hfil
             \begin{minipage}{58mm}
                    \includegraphics[width = \linewidth]{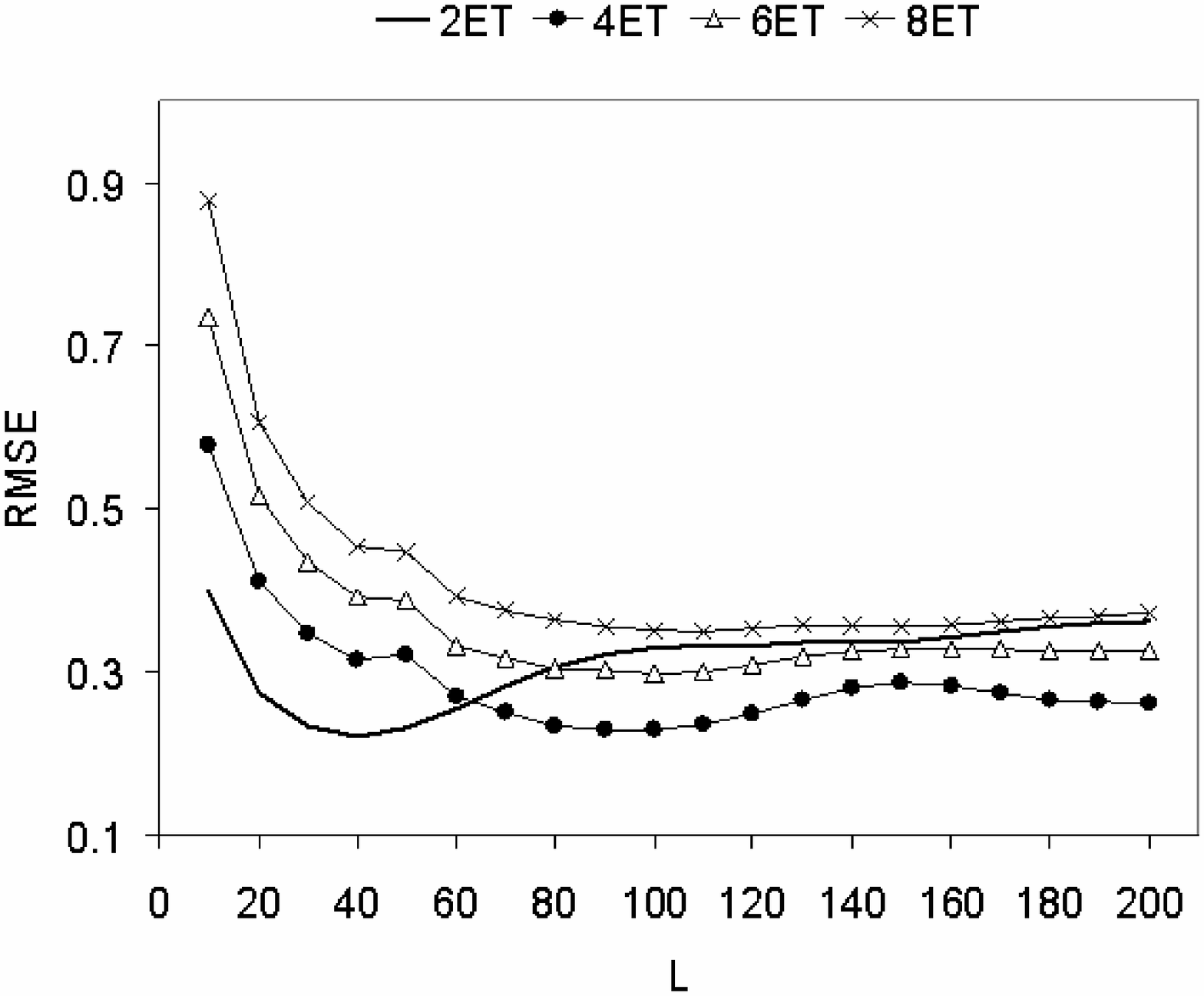}
                    \vspace{-3mm}\caption[t]{RMSE of reconstruction: t.s.~\eqref{eq:mod2}, $\sigma=1$}
                    \label{fig:cos2cos10MSE}
             \end{minipage}
             \end{figure}
\end{center}

\smallskip\noindent
\textbf{The case of complex-form modulation}
Let us consider the case where the modulated signal is not a signal of finite rank.
Let $N=399$ and $f_n=s_n+r_n$, where
\begin{gather}
\label{eq:mod2}
  s_n=A(n)\cos(2\pi n/20),\ r_n=\sigma \varepsilon_n,
\end{gather}
where $A(n)=\cos(2\pi n^2/10^5)$.
Figures \ref{fig:cos2cos01series} and \ref{fig:cos2cos10series} show the initial time series with different levels of noise
and Figures \ref{fig:cos2cos01MSE} and \ref{fig:cos2cos10MSE} contain the errors of reconstruction by 2, 4, 6, and 8 leading eigentriples.
The choice of the window length for signal reconstruction is not crucial for low levels of noise, since we are able to achieve a good accuracy
by choosing eigentriples for reconstruction properly (the larger the window length $L\le N/2$, the larger the number of eigentriples chosen
for reconstruction).

%\vspace{-1cm}

\vspace{-1.5cm}
Fig.~\ref{fig:cos2cos10MSE} shows that for a high level of noise the choice of the window length close to $N/2$ does not
provide the best accuracy.
In particular, using $L=200$ with reconstruction by four leading eigentriples is slightly worse than using $L=40$ (two basic periods)
and two leading eigentriples. However, if we take, say, $L=80$, then reconstruction by two leading eigentriples
is less accurate.
Thus, the choice of $L\sim N/2$ is quite appropriate and can be improved only with the help of
additional information (like the value of period) about the analyzed time series.

Note that in this subsection we considered the choice of window lengths for signal reconstruction.
Estimation of the basic frequency (e.g., by ESPRIT)
is a separate problem, which requires special statements and approaches to its solution.

\subsection{The problem of mixing}
One of the problems that SSA encounters is a possible lack of strong separability \cite{Golyandina.etal2001} under conditions
of weak separability. This problem is caused by equal singular values in SVDs of trajectory matrices of
the signal and the residual.
%Note that the previous examples from this section can be partly explained by mixing.

Let us consider some deterministic slowly varying component (i.e., a trend) as a signal.
The signal does not necessarily have a finite rank. More likely, it can be approximated
by a time series of finite rank $r_{\mathrm{appr}}$. Let $Q$ be the share of $r_{\mathrm{appr}}$ leading squared singular values
in the SVD of the trajectory matrix of the trend.
Let us denote by $\lambda_{\mathrm{max}}^{\mathrm{resid}}$ the maximal singular value generated by
the trajectory matrix of the residual and by $\lambda_{\mathrm{min}}^{\mathrm{trend}}$
the $r_{\mathrm{appr}}$-th singular value of the trajectory matrix of the trend.

The case of approximate separability corresponds to large $Q$ under the condition
$\lambda_{\mathrm{max}}^{\mathrm{resid}}<\lambda_{\mathrm{min}}^{\mathrm{trend}}$.
If $Q$ is fixed, then, in general, the larger the window length, the larger the rank
of the approximating time series and the smaller is $\lambda_{\mathrm{min}}^{\mathrm{trend}}$.
This observation can lead to the optimal window lengths considerably smaller than $N/2$.

\begin{center}
             \begin{figure}[h!]
             \begin{minipage}{58mm}
                    \includegraphics[width = \linewidth]{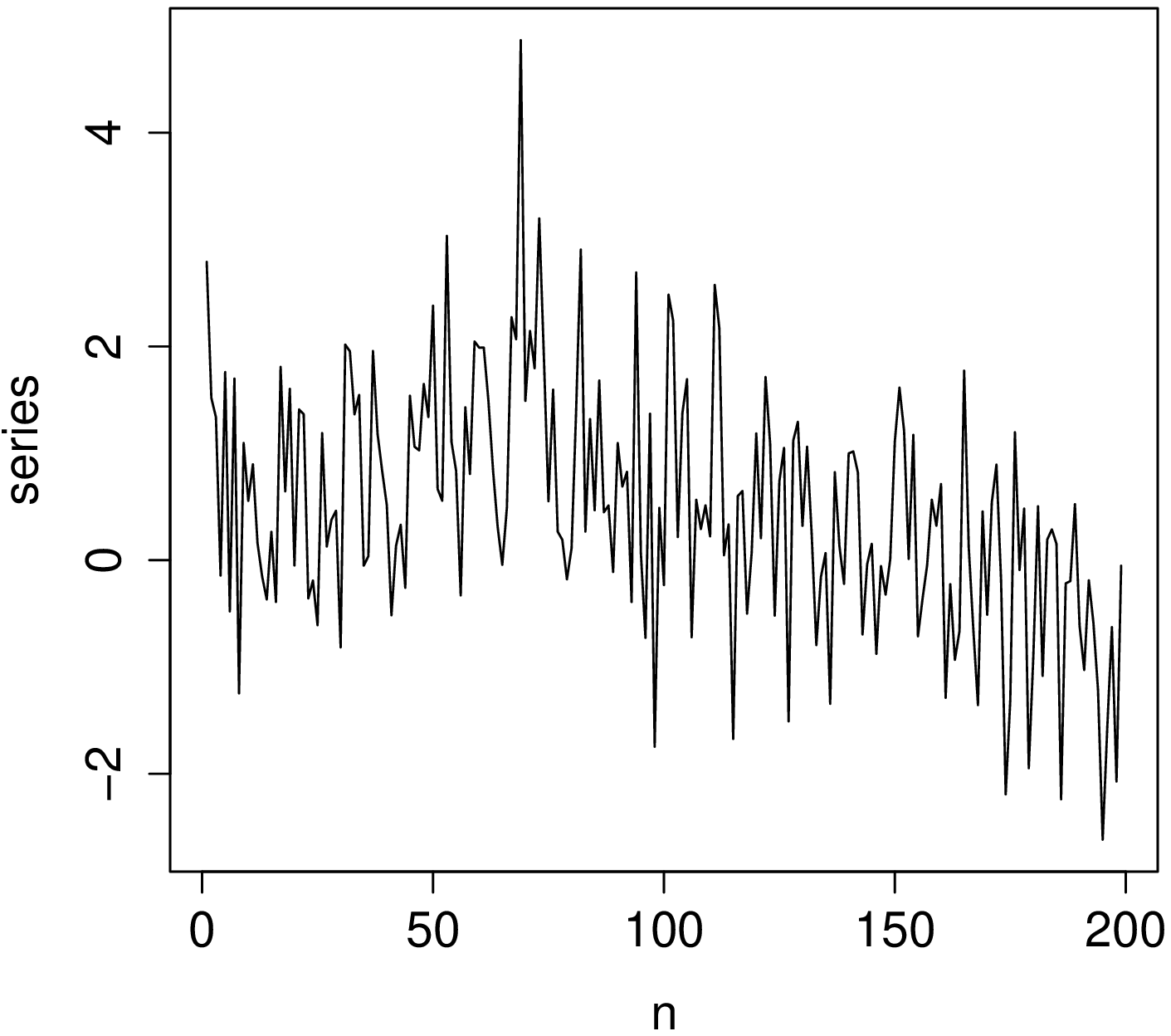}
                    \vspace{-5mm}\caption[t]{Initial time series: t.s.~\eqref{eq:modmix}}
                    \label{fig:cos2noise10cos05_series}
             \end{minipage}
             \hfil
             \begin{minipage}{58mm}
                    \includegraphics[width = \linewidth]{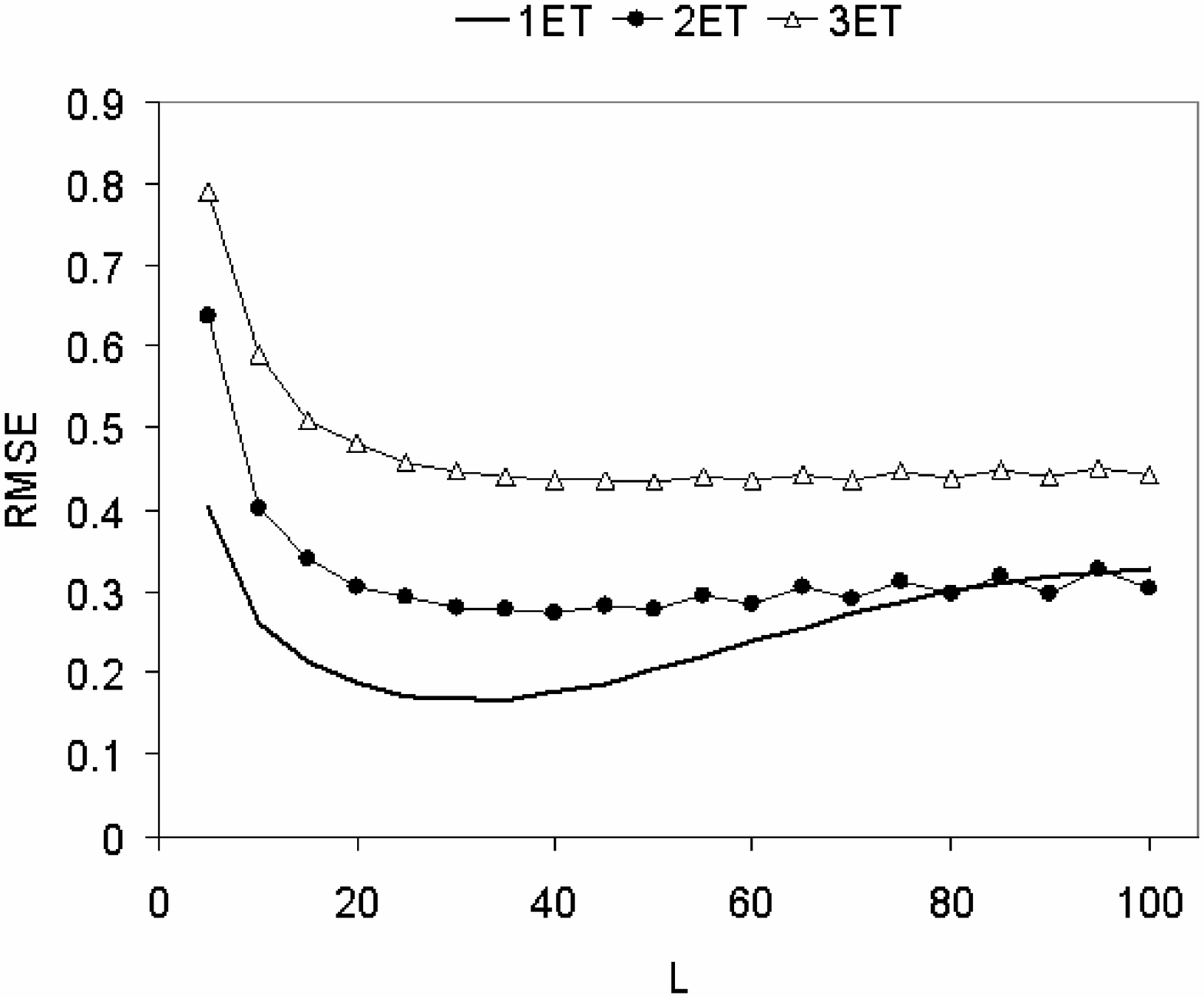}
                    \vspace{-5mm}\caption[t]{RMSE of reconstruction: t.s.~\eqref{eq:modmix}}
                    \label{fig:cos2noise10cos05}
             \end{minipage}
             \end{figure}
\end{center}

We consider the example
\be
\label{eq:modmix}
  s_n=\cos(2\pi n^2/10^5),\ r_n=\varepsilon_n+0.5\cos(2\pi n/10),
\ee
$N=199$ (Fig. \ref{fig:cos2noise10cos05_series}). It is easy to check numerically that for $L=100$ we can get $r_{\mathrm{appr}}=2$ with
$Q=99.5\%=92\%+7.5\%$ (the singular values are equal to 74 and 21), whereas for $L=30$,  $r_{\mathrm{appr}}=1$ with $Q=98\%$
(the singular value is 41).
However, the residual produces the maximal singular value equal approximately to 25 for $L=100$ and to 15 for $L=30$.
This means that we have no strong separability for $L=100$ with $Q=99.5\%$. Thus, we can reach $Q=92.5\%$ for
$L=100$ and $Q=98\%$ for $L=30$ (certainly, these are just rough measures)
to satisfy $\lambda_{\mathrm{max}}^{\mathrm{resid}}<\lambda_{\mathrm{min}}^{\mathrm{trend}}$, that is
confirmed by Fig. \ref{fig:cos2noise10cos05}.
Note that after extracting the trend with $L_1=30$ the periodicity can be extracted from the residual
with a window length close to $L_2=100$. In \cite{Golyandina.etal2001} this technique
is called Sequential SSA.

\vspace{-1cm}
\section{SSA processing of stationary time series}
\label{sec:stationary}
There are special recommendations concerning the choice of parameters for stationary time series.
%(see usually, the stationarity means approximate constancy of average t.s. values and of average squared t.s.
%values over long time intervals).
It is traditionally recommended to perform the centering procedure for the stationary time series
before processing (i.e., to subtract the average over the time series)
and then to use the Toeplitz autocovariance matrix $\widetilde{\mathbf{C}}$ with entries
\bea
 \widetilde c_{ij} = \frac{1}{N-|i-j|}\sum_{m=0}^{N-|i-j|-1}f_m f_{m+|i-j|}
\eea
instead of $\mathbf{C}=\mathbf{X}\mathbf{X}^\mathrm{T}$
at the decomposition stage (see \cite{Golyandina.etal2001} for details of the Toeplitz SSA algorithm).
Let us remark that using $\mathbf{C}$ to obtain the SVD of the trajectory matrix is sometimes called `BK' following \cite{Broomhead.King1986},
while using $\widetilde{\mathbf{C}}$ to get the eigenvectors is initiated by the spectral analysis and is called `VG' following
\cite{Vautard.Ghil1989}. The use of $\widetilde{\mathbf{C}}$ does not provide us with the SVD of the trajectory matrix and
therefore with the SVD optimality.

The papers related to the SSA analysis of climate time series (e.g. \cite{Ghil.etal2002}) consider the Toeplitz SSA
as the main version and state that the Basic and Toeplitz versions only slightly differ.
Our investigation shows that Toeplitz SSA provides more stable SSA results (reconstruction, forecast, estimates).
However, these results can be inadequate and can have a considerable bias if the time series we analyze is not stationary.
It seems that using the Toeplitz version of the SSA algorithm is unsafe if
the time series contains a trend or oscillations with increasing or decreasing amplitudes.

Here we can apply a well-known principle: if the method assumes a model, then it gives more precise results
when the model is valid; otherwise, the method can produce completely wrong results.

Centering the time series is a less risky procedure than using $\widetilde{\mathbf{C}}$
and can either slightly improve the SSA results or worsen them.
Let us note that centering usually increases the rank of the signal making the signal
structure more complicated.

Let us recall that besides SSA with centering as preprocessing there are the so called ``Single centering SSA'' \cite{Golyandina.etal2001},
which is appropriate for time series with a constant trend, and also the so called ``Double centering SSA'' \cite{Golyandina.etal2001}, which
works well for time series with linear trends.

In the following subsections we demonstrate the examples of application of the centering procedure and
the Toeplitz SSA algorithm to non-stationary time series.

\subsection{Centering as preprocessing}
Let $N=199$ and $f_n=s_n+r_n$ with
\be
\label{eq:centr}
  s_n=1.005^n,\ r_n=\varepsilon_n,
\ee
see Fig.~\ref{fig:exp_series}.

The rank of the exponential series is 1 and therefore for its extraction
we should choose just one eigentriple.
After centering this time series, we obtain a new one $G_N=(g_0,\ldots,g_{N-1})$ with $g_n=1.005^n - c +\varepsilon_n$.
Therefore, we should choose two eigentriples to extract $1.005^n - c$, i.e., we artificially create a more complex structure of the signal.
The results of simulations (see Fig.~\ref{fig:exp_MSE}, $L$ changes from 5 to 100 with increment 5)
confirm that the thickening of the signal structure ends in the increase of errors.

Practically, the centering procedure can increase the errors of reconstruction even of undamped sinusoids
in short time series if the time series length is not divisible by the sine-wave period.
The explanation is similar to that for the exponential time series: if the time series length is not divisible by the sine-wave period, then
the average over the time series is not equal to 0. Therefore, after subtracting this average
we transform the signal of rank 2 to a signal of rank 3. For long time series the effect of the rank increase is
diminished because the time series average is almost zero. However, in such a case centering has no sense.

\vspace{-5mm}
\begin{center}
             \begin{figure}[h!]
                    \includegraphics[width = 58mm]{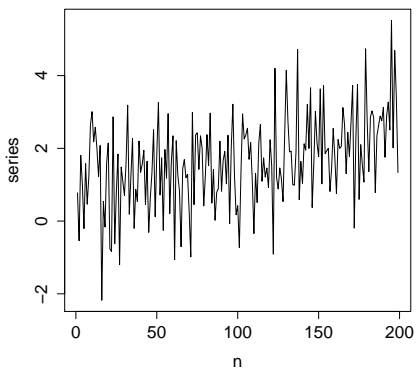}
                    \vspace{-3mm}\caption[t]{Initial time series: t.s.~\eqref{eq:centr}}
                    \label{fig:exp_series}
             \end{figure}
              \begin{figure}[h!]
                    \includegraphics[width = 58mm]{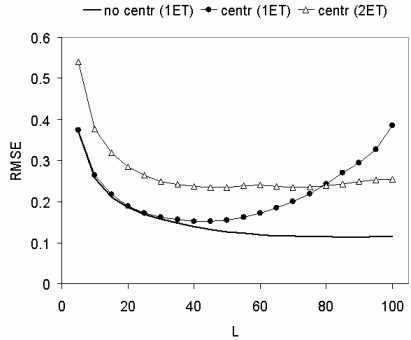}
                    \vspace{-3mm}\caption[t]{RMSE of reconstruction: t.s.~\eqref{eq:centr}, reconstruction by ET1 with no centering
                    and by ET1--2 with the preliminary centering}
                    \label{fig:exp_MSE}
             \end{figure}
\end{center}

\vspace{-5mm}
\begin{remark}[About filling in the missing data] \rm
Let us mention that the method of imputation of missing values introduced in \cite{Golyandina.Osipov2007} does not imply centering
(although centering can be used).
In the method from \cite{Kondrashov.Ghil2006} the centering procedure is essential, since the missing values are replaced with
zeros at the first iteration. It seems that this peculiarity is caused by the fact
that the first applications of SSA have been oriented at stationary time series.
Certainly, to generalize the method proposed in \cite{Kondrashov.Ghil2006}, we can fill in
the missing values using the average over the whole time series or use an interpolation method based on neighboring non-missing values.
\end{remark}

\begin{remark}[About eigenvalues] \rm
Eigenvalues play an important role in the analysis of stationary time series (see, for example, papers devoted to Monte Carlo SSA
for detecting a signal in red noise \cite{Allen.Smith1996}).
For stationary time series, the share of eigentriples corresponding to the signal represents the contribution of the corresponding
component.
However, in the general case of arbitrary time series, there is little point in the eigentriples share, since it depends on
absolute values of the times series: in Basic SSA, the distribution of eigenvalues (mostly, leading) depends on the constant compound of
the time series while the structure of the time series should not depend on addition or substraction of a constant.
In particular, if the leading eigentriple takes 99.9\%, this does not mean that it is enough to take
only one eigentriple to approximate the time series with high accuracy.
Certainly, we can use information on eigenvalues to identify pairs of eigentriples generated by sinusoids
(one sinusoid produces two close or equal eigenvalues). However, the same identification can be done
by more powerful tools.

\end{remark}

\subsection{Toeplitz SSA}

Let us demonstrate the consequences of improper use of Toeplitz SSA.
First, application of Toeplitz SSA to non-stationary signals of finite rank $r$
generally increases the number of nonzero eigenvalues from $r$ to the maximal possible value
equal to $\min(L,K)$, that is, the structure can be lost. This means that the number of eigentriples
required for accurate reconstruction increases. Also, the constructed approximation can have a
wrong structure and, for example, lead to a wrong forecast.
\begin{center}
             \begin{figure}[h!]
             \begin{minipage}{70mm}
                    \includegraphics[width = \linewidth]{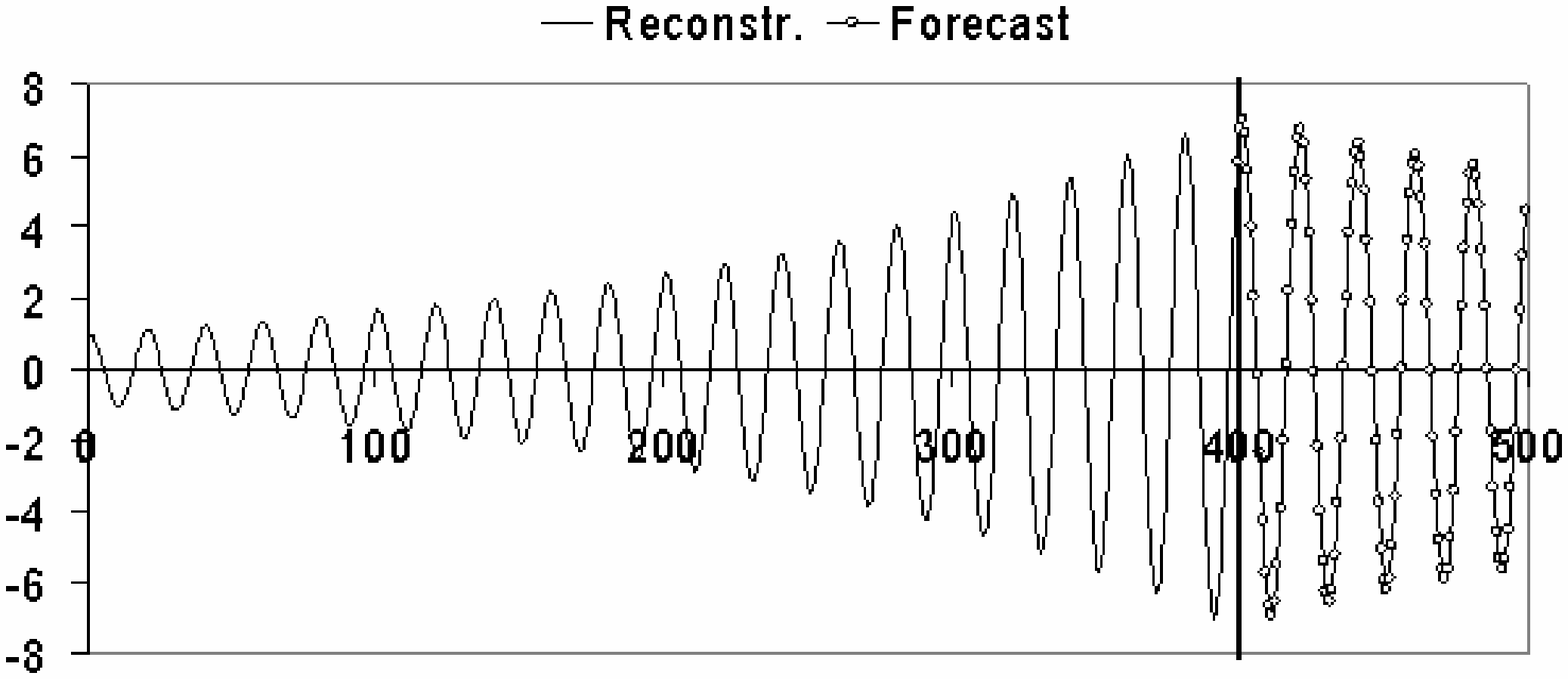}
                    \vspace{-3mm}\caption[t]{Toeplitz SSA forecast for $f_n=1.005^n\cos(2\pi n/20)$, $N=399$, $L=200$, ET1--14}
                    \label{fig:toeplitz_cos}
             \end{minipage}
             \hfil
             \begin{minipage}{70mm}
                    \includegraphics[width = \linewidth]{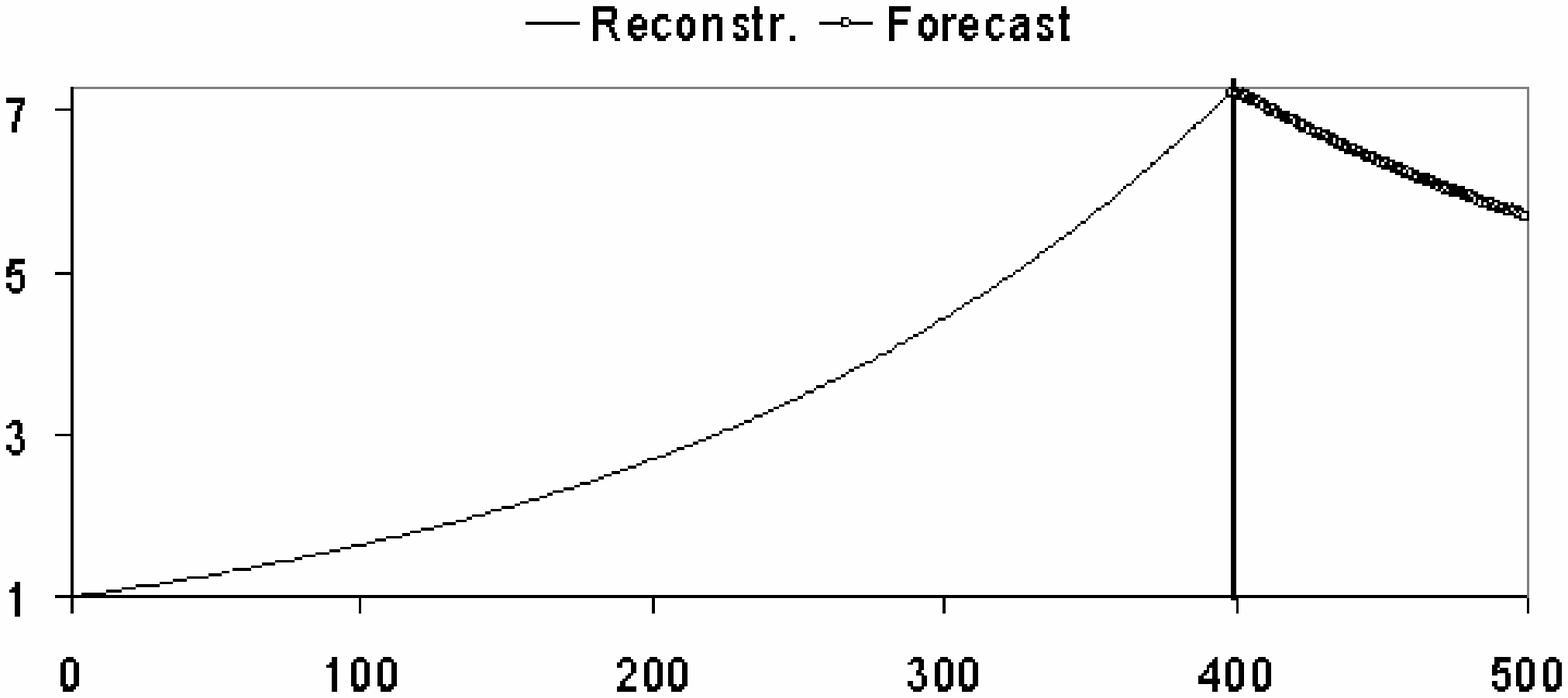}
                    \vspace{-3mm}\caption[t]{Toeplitz SSA forecast for $f_n=1.005^n$, $N=399$, $L=200$, ET1--14}
                   \label{fig:toeplitz_exp}
             \end{minipage}
             \end{figure}
\end{center}

\vspace{-6mm}
We consider two examples of time series of finite rank,
$f_n=1.005^n\cos(2\pi n/20)$ (rank 2) and $f_n=1.005^n$ (rank~1).
To get an accurate approximation of the last time series points, we perform Toeplitz-SSA reconstruction with $L=200$
based on ET1--14 (ET1--2 for the first example and ET1 for the second one do not provide good accuracy).
Figures \ref{fig:toeplitz_cos} and \ref{fig:toeplitz_exp}
demonstrate that the forecast based on the chosen components is inadequate (moreover, this conclusion
does not depend on the chosen window length $L$ and number of components).

One can see that the Toeplitz-SSA forecast is completely wrong while the Basic SSA forecast
of finite-rank time series is precise.

\section{SVD-origins of SSA and the choice of SSA parameters}
\label{sec:origins}

The key step in SSA is the Singular Value Decomposition (SVD) of the trajectory matrix. The SVD is used for solving different problems,
including statistical methods in data analysis. Therefore, the logic of these procedures can also be extended to SSA.
Let us briefly describe several origins of SVD-related ideas and views on the SSA-parameter choice generated by them.
We do not aim to review the literature on SVD but instead just emphasize the relation between origins and the methodology of SSA.

\smallskip\noindent
\textbf{Principal Component Analysis (PCA)\ } This origin is characterized by different kinds of manipulations with variables
and with cases in data, i.e., with rows and columns of the trajectory matrix. In particular,
the conventional manipulations are centering and standardization of variables.
This induces different views on
eigenvectors and factor vectors (vectors of principal components), where the latter is interpreted as components of the original time series
(maybe, just a little shorter than the original time series).
In addition, greater attention to eigenvalues' contribution is transferred from PCA to SSA.
This attitude to the trajectory matrix is appropriate when we apply Single centering SSA and when
the number of rows ($L$) is fixed and is smaller than the number of columns ($K$).
In the general case, the structure of the trajectory matrix does not depend on the transposition of the trajectory matrix,
the interpretation of eigenvalues is not so important and, in particular, Double centering is often more natural than Single one.

\smallskip\noindent
\textbf{Hankel rank-deficient matrices\ } The relation between such matrices and time series governed by linear recurrent formulas has long been known
  (see e.g. \cite{Gantmacher1959}). This technique allows us to analyze noisy time series governed by LRFs. The main
  application of this idea  refers to the signal processing with its approaches to the choice of the method's parameters
  and consists in the analysis of a noisy sum of damped/undamped cisoids
  and the estimation of their parameters (mostly, frequencies). However, this approach can be applied to parameter estimation of
  arbitrary signals governed by LRFs.

\smallskip\noindent
\textbf{Spectral analysis \cite{Ghil.Vautard1991}\ } The name of this approach is close to
 Singular Spectrum Analysis. However, the name `Spectral analysis' means the analysis of stationary time series
 and their frequency characteristics. On the other hand, Singular Spectrum is related to the spectrum of linear operators,
 i.e. singular values of trajectory matrices in the case of SSA. Thus, the analysis of Singular Spectrum does not imply stationarity.
 As has been stated in \cite{Elsner.Tsonis1996}, the name SSA does not reflect the multifaceted entity of SSA and is traditionally used.
 Peculiarity of this origin consists in centering the time series before processing and subsequent application of the Toeplitz version of SSA.
 As was demonstrated in Section~\ref{sec:stationary}, if we apply this technique to non-stationary time series,
 it is likely that we obtain either imprecise or even  meaningless results.
 Within the framework of Spectral analysis approach, special attention is paid to
 the red noise (autoregression of order 1) and testing the null hypothesis of the absence of a signal in red noise.
 In some sense, this corresponds to the  problem with weak signal and strong noise (in contrast to the previous origin).

\smallskip\noindent
\textbf{Karhunen-Lo\`{e}ve expansion \cite{Basilevsky.Hum1979}\ } This expansion is conventionally used in the theory of stochastic processes and
  originally assumes zero expectation of elements of the considered stochastic process (or subtracting the known averages).
  In \cite{Basilevsky.Hum1979},  estimation of  the process average is performed by using the moving average.
  Then the errors from this estimation are included in the centered stochastic process and this allows one to apply
  the method to processes with trends. Thus, the algorithm appears to be very close to PCA and in fact coincides with Single centering SSA.
  It seems that this origin is
  used by researchers who are well familiar with the stochastic process techniques and therefore the association with KL expansion
  helps them to understand the SSA method.

\smallskip\noindent
\textbf{Dynamical systems \cite{Broomhead.King1986,Fraedrich1986}\ } This origin is related to special
problems in the theory of dynamical systems
with a specific approach to the choice of parameters.
However, the contribution of these papers is considerable, since the described algorithm served
as an origin of SSA ideas in several applied areas.

\section{Conclusion}

In this paper we have considered the SSA-related methods from a unified point of view. The approach to the
investigation of these methods was formulated for the `signal + residual' time series.
This enabled us to show the similarity and the specifics of the problems which can be solved by the considered methods.

%In particular, we considered decomposed residuals into several components,
%which are allowed natural interpretation, and classified the problems by the type of the use of the SVD
%of the trajectory matrix.
The accuracy of the methods and the choice of parameters were studied.
In the paper we emphasized on the internal mechanism of error origins.
The understanding of this internal mechanism together with computer simulations for several typical examples
enabled us to formulate the recommendations on the optimal choice of the window length $L$, which is the main parameter in SSA.
In particular, the choice of $L$ close to one-half of the time series length was approved as appropriate in most cases.
Classes of time series for which this choice should be corrected were indicated.

\section*{Acknowledgments}
The author is thankful to the anonymous referees, the editors and also to Andrey Pepelyshev for their useful comments and suggestions
which helped to improve the paper.

\bibliography{ssa_esprit}

\begin{thebibliography}{}
\ifx \url   \undefined \def \url#1{#1}   \fi

\bibitem{Allen.Smith1996}
\textsc{Allen, M.} \textsc{and} \textsc{Smith, L.} (1996).
\newblock {Monte Carlo SSA}: Detecting irregular oscillations in the presence
  of colored noise.
\newblock \emph{Journal of Climate\/}~\textbf{9},~12, 3373--3404.

\bibitem{Badeau.etal2008}
\textsc{Badeau, R.}, \textsc{Richard, G.}, \textsc{and} \textsc{David, B.}
  (2008).
\newblock Performance of {ESPRIT} for estimating mixtures of complex
  exponentials modulated by polynomials.
\newblock \emph{IEEE Transactions on Signal Processing\/}~\textbf{56},~2,
  492--504.

\bibitem{Barkhuijsen.etal1987}
\textsc{Barkhuijsen, H.}, \textsc{de~Beer, R.}, \textsc{and} \textsc{van
  Ormondt, D.} (1987).
\newblock Improved algorithm for noniterative time-domain model fitting to
  exponentially damped magnetic resonance signals.
\newblock \emph{J. Magn. Reson.\/}~\emph{73}, 553--557.

\bibitem{Basilevsky.Hum1979}
\textsc{Basilevsky, A.} \textsc{and} \textsc{Hum, D. P.~J.} (1979).
\newblock Karhunen-{L}o\'eve analysis of historical time series with an
  application to plantation births in {Jamaica}.
\newblock \emph{{J. Am. Stat. Assoc.}\/}~\emph{74}, 284--290.

\bibitem{Bjorck1996}
\textsc{Bj\"orck, A.} (1996).
\newblock \emph{Numerical Methods for Least Squares Problems}.
\newblock SIAM.

\bibitem{Broomhead.King1986}
\textsc{Broomhead, D.~S.} \textsc{and} \textsc{King, G.~P.} (1986).
\newblock Extracting qualitative dynamics from experimental data.
\newblock \emph{Physica D\/}~\emph{20}, 217--236.

\bibitem{Cadzow1988}
\textsc{Cadzow, J.~A.} (1988).
\newblock Signal enhancement: a composite property mapping algorithm.
\newblock \emph{IEEE Transactions on Acoustics, Speech, and Signal
  Processing\/}~\textbf{36},~1, 49--62.

\bibitem{Groen1996}
\textsc{de~Groen, P.} (1996).
\newblock An introduction to total least squares.
\newblock \emph{Nieuw Archief voor Wiskunde\/}~\emph{14}, 237--253.

\bibitem{Djermoune.Tomczak2009}
\textsc{Djermoune, E.-H.} \textsc{and} \textsc{Tomczak, M.} (2009).
\newblock Perturbation analysis of subspace-based methods in estimating a
  damped complex exponential.
\newblock \emph{IEEE Transactions on Signal Processing\/}~\textbf{57},~11,
  4558--4563.

\bibitem{Elsner.Tsonis1996}
\textsc{Elsner, J.~B.} \textsc{and} \textsc{Tsonis, A.~A.} (1996).
\newblock \emph{{S}ingular {S}pectrum {A}nalysis: A New Tool in Time Series
  Analysis}.
\newblock Plenum.

\bibitem{Fraedrich1986}
\textsc{Fraedrich, K.} (1986).
\newblock Estimating dimensions of weather and climate attractors.
\newblock \emph{J. Atmos. Sci.\/}~\emph{43}, 419--432.

\bibitem{Gantmacher1959}
\textsc{Gantmacher, F.~R.} (1959).
\newblock \emph{The Theory of Matrices}.
\newblock Chelsea Publishing Company, New York 68.

\bibitem{Ghil.etal2002}
\textsc{Ghil, M.}, \textsc{Allen, R.~M.}, \textsc{Dettinger, M.~D.},
  \textsc{Ide, K.}, \textsc{Kondrashov, D.}, \textsc{Mann, M.~E.},
  \textsc{Robertson, A.}, \textsc{Saunders, A.}, \textsc{Tian, Y.},
  \textsc{Varadi, F.}, \textsc{and} \textsc{Yiou, P.} (2002).
\newblock Advanced spectral methods for climatic time series.
\newblock \emph{Rev. Geophys.\/}~\textbf{40},~1, 1--41.

\bibitem{Ghil.Vautard1991}
\textsc{{Ghil}, M.} \textsc{and} \textsc{{Vautard}, R.} (1991).
\newblock {Interdecadal oscillations and the warming trend in global
  temperature time series}.
\newblock \emph{Nature\/}~\emph{350}, 324--327.

\bibitem{Golyandina.etal2001}
\textsc{Golyandina, N.}, \textsc{Nekrutkin, V.}, \textsc{and}
  \textsc{Zhigljavsky, A.} (2001).
\newblock \emph{Analysis of Time Series Structure: {SSA} and Related
  Techniques}.
\newblock Chapman~\&~Hall/CRC.

\bibitem{Golyandina.Osipov2007}
\textsc{Golyandina, N.} \textsc{and} \textsc{Osipov, E.} (2007).
\newblock The ``{Caterpillar}''-{SSA} method for analysis of time series with
  missing values.
\newblock \emph{J. Stat. Plan. Infer.\/}~\textbf{137},~8, 2642--2653.

\bibitem{Golyandina.Vlassieva2009}
\textsc{Golyandina, N.} \textsc{and} \textsc{Vlassieva, E.} (2009).
\newblock First-order {SSA}-errors for long time series: model examples of
  simple noisy signals.
\newblock In \emph{Proceedings of the 6th St.Petersburg Workshop on Simulation
  Vol.1, June 28-July 4, 2009, St. Petersburg}. St.Petersburg State University,
  314--319.

\bibitem{Hall1998}
\textsc{Hall, M.~J.} (1998).
\newblock \emph{Combinatorial theory}.
\newblock Wiley, New York.

\bibitem{Kondrashov.Ghil2006}
\textsc{Kondrashov, D.} \textsc{and} \textsc{Ghil, M.} (2006).
\newblock Spatio-temporal filling of missing points in geophysical data sets.
\newblock \emph{Nonlinear Processes in Geophysics\/}~\textbf{13},~2, 151--159.

\bibitem{Korobeynikov2009}
\textsc{Korobeynikov, A.} (2010).
\newblock Computation- and space-efficient implementation of {SSA}.
\newblock \emph{Statistics and Its Interface\/}~\emph{3}, 357--368.

\bibitem{Krim.etal1992}
\textsc{Krim, H.}, \textsc{Forster, P.}, \textsc{and} \textsc{Proakis, J.~G.}
  (1992).
\newblock Operator approach to performance analysis of root-{MUSIC} and
  root-min-norm.
\newblock \emph{IEEE Transactions on Signal Processing\/}~\textbf{40},~7
  (July), 1687--1696.

\bibitem{Kumaresan.Tufts1982}
\textsc{Kumaresan, R.} \textsc{and} \textsc{Tufts, D.} (1982).
\newblock Estimating the parameters of exponentially damped sinusoids and
  pole-zero modeling in noise.
\newblock \emph{IEEE Transactions on Acoustics, Speech, and Signal
  Processing\/}~\textbf{30},~6, 833--840.

\bibitem{Kumaresan.Tufts1980}
\textsc{Kumaresan, R.} \textsc{and} \textsc{Tufts, D.~W.} (1980).
\newblock Data-adaptive principal component signal processing.
\newblock In \emph{Proc. of {IEEE} Conference On Decision and Control}.
  Albuquerque, 949--954.

\bibitem{Kumaresan.Tufts1983}
\textsc{Kumaresan, R.} \textsc{and} \textsc{Tufts, D.~W.} (1983).
\newblock Estimating the angles of arrival of multiple plane waves.
\newblock \emph{IEEE Transactions on Aerospace and Electronic
  Systems\/}~\textbf{AES-19},~1, 134--139.

\bibitem{Kung.etal1983}
\textsc{Kung, S.~Y.}, \textsc{Arun, K.~S.}, \textsc{and} \textsc{Rao, D. V.~B.}
  (1983).
\newblock State-space and singular-value decomposition-based approximation
  methods for the harmonic retrieval problem.
\newblock \emph{J. Opt. Soc. Am.\/}~\textbf{73},~12, 1799--1811.

\bibitem{Li.etal1993}
\textsc{Li, F.}, \textsc{Liu, H.}, \textsc{and} \textsc{Vaccaro, R.~J.} (1993).
\newblock Performance analysis for {DOA} estimation algorithms: unification,
  simplification, and observations.
\newblock \emph{IEEE Transactions on Aerospace and Electronic
  Systems\/}~\textbf{29},~4, 1170--1184.

\bibitem{Nekrutkin2010}
\textsc{Nekrutkin, V.} (2010).
\newblock Perturbation expansions of signal subspaces for long signals.
\newblock \emph{Statistics and Its Interface\/}~\emph{3}, 297--319.

\bibitem{Pakula1987}
\textsc{Pakula, L.} (1987).
\newblock Asymptotic zero distribution of orthogonal polynomials in sinusoidal
  frequency estimation.
\newblock \emph{IEEE Trans. Inf. Theor.\/}~\textbf{33},~4, 569--576.

\bibitem{Rife.Boorstyn1974}
\textsc{Rife, D.} \textsc{and} \textsc{Boorstyn, R.} (1974).
\newblock Single tone parameter estimation from discrete-time observations.
\newblock \emph{Information Theory, IEEE Transactions on\/}~\textbf{20},~5,
  591--598.

\bibitem{Roy.Kailath1989}
\textsc{{Roy}, R.} \textsc{and} \textsc{{Kailath}, T.} (1989).
\newblock {ESPRIT: estimation of signal parameters via rotational invariance
  techniques}.
\newblock \emph{IEEE Trans. Acoust.\/}~\emph{37}, 984--995.

\bibitem{Stoica.etal1997}
\textsc{Stoica, P.}, \textsc{Jakobsson, A.}, \textsc{and} \textsc{Li, J.}
  (1997).
\newblock Cisoid parameter estimation in the colored noise case: asymptotic
  {C}ramer-{R}ao bound, maximum likelihood, and nonlinear least-squares.
\newblock \emph{{IEEE Transactions on Signal Processing}\/}~\textbf{45},~8
  (Aug.), 2048--2059.

\bibitem{Stoica.Moses1997}
\textsc{Stoica, P.} \textsc{and} \textsc{Moses, R.} (1997).
\newblock \emph{Introduction to Spectral Analysis}.
\newblock Prentice Hall.

\bibitem{Tufts.Kumaresan1982b}
\textsc{Tufts, D.~W.} \textsc{and} \textsc{Kumaresan, R.} (1982).
\newblock Estimation of frequencies of multiple sinusoids: Making linear
  prediction perform like maximum likelihood.
\newblock \emph{Proceedings of the IEEE\/}~\textbf{70},~9 (Sept.), 975--989.

\bibitem{Usevich2010}
\textsc{Usevich, K.} (2010).
\newblock On signal and extraneous roots in {Singular Spectrum Analysis}.
\newblock \emph{Statistics and Its Interface\/}~\emph{3}, 281--295.

\bibitem{VanHuffel.etal1994}
\textsc{Van~Huffel, S.}, \textsc{Chen, H.}, \textsc{Decanniere, C.},
  \textsc{and} \textsc{van Hecke, P.} (1994).
\newblock Algorithm for time-domain {NMR} data fitting based on total least
  squares.
\newblock \emph{J. Magn. Reson. Ser. A\/}~\emph{110}, 228--237.

\bibitem{Vautard.Ghil1989}
\textsc{Vautard, M.} \textsc{and} \textsc{Ghil, M.} (1989).
\newblock Singular spectrum analysis in nonlinear dynamics, with applications
  to paleoclimatic time series.
\newblock \emph{Physica D\/}~\emph{35}, 395--424.

\end{thebibliography}

\end{document}